\documentclass[fleqn,usenatbib,useAMS]{mnras}

\usepackage{graphicx}	% Including figure files
\usepackage{amsmath}	% Advanced maths commands
\usepackage{amssymb}	% Extra maths symbols
\usepackage{multicol}        % Multi-column entries in tables
\usepackage{bm}		% Bold maths symbols, including upright Greek
\usepackage{pdflscape}	% Landscape pages

\usepackage[T1]{fontenc}
\usepackage{ae,aecompl}

\usepackage{newtxtext,newtxmath}

\def\eq#1{{Eq.~(\ref{#1})}}

\title[Synergies of 21 cm and sub-mm]{Synergizing 21 cm and sub-millimetre surveys during reionization: new empirical insights}

\author[]{Hamsa Padmanabhan\thanks{hamsa.padmanabhan@unige.ch}
\\
D\'epartement de Physique Th\'eorique, Universit\'e de Gen\`eve,
24 quai Ernest-Ansermet, CH 1211 Gen\`eve 4, Switzerland
\\
}

\date{\today}

\pubyear{2022}

\begin{document}
\label{firstpage}
\pagerange{\pageref{firstpage}--\pageref{lastpage}}
\maketitle

\begin{abstract}
We use the latest results from Atacama Large Millimetre/submillimetre Array (ALMA) surveys targeting the ionized carbon [CII] 158 $\mu$m and oxygen [OIII] 88 $\mu$m lines, in combination with data-driven predictions for the evolution of neutral hydrogen (HI), to  illustrate the prospects for intensity mapping cross-correlations between 21 cm and submillimetre surveys over $z \sim 5-7$. We work with a dataset including the ALPINE and REBELS surveys for [CII] over $z \sim 4.5-7$, and ALMA  [OIII] detections over $z \sim 6-9$. The resultant evolution of the [CII] luminosity - halo mass relation is well described by a double power law at high redshifts, with the best-fitting parameters in good agreement with the results of simulations. The data favour secure detections of the auto-power spectrum of [CII] at all redshifts with an enhanced Fred Young Submillimetre Telescope (FYST)-like configuration. Such an experiment, along with the Murchison Widefield Array (MWA) will be able to measure the 21 cm - [CII] cross-correlation  power with a signal-to-noise ratio of a few tens to a few hundreds. We find that a balloon-borne experiment improving upon the Experiment for Cryogenic Large-Aperture Intensity Mapping (EXCLAIM)  should be able to detect the 21 cm - [OIII] cross-correlation with the MWA and the Square Kilometre Array (SKA)-LOW out to $z \sim 7$. Our  results have implications for constraining  the evolution of luminous sources during the mid-to-end stages of reionization.
\end{abstract}

\begin{keywords}
dark ages, reionization, first stars  -- galaxies: high redshift -- submillimetre: ISM
\end{keywords}

\begingroup
\let\clearpage\relax
\endgroup
\newpage

\section{Introduction}
The epoch of reionization ($z \sim 6-12$), when the newly formed luminous sources ionized the hydrogen in the intergalactic medium (IGM), is a key period in the history of evolution of the Universe \citep[for a review, see, e.g.,][]{barkana2001}. Mapping the redshifted 21 cm emission by neutral hydrogen (HI) in the  IGM will offer a three-dimensional view into the earliest galaxies during this period, whose complementary Lyman-$\alpha$ line is saturated by absorption in the IGM \citep{fan2006, ota2017}.  Many experiments are currently functioning (or in the final stages of commissioning) to produce such maps of the large-scale 21 cm signal and measure its clustering during the period of reionization,  such as the the Low Frequency Array \citep[LOFAR;][]{vanhaarlem2013, pober2014}, the Hydrogen Epoch of Reionization Array \citep[HERA,][]{hera2022, pober2014, deboer2017}, and the Murchison Widefield Array \citep[MWA;][]{bowman2013, tingay2013}, eventually leading up to the Square Kilometre Array\footnote{https://www.skao.int/en/explore/telescopes/ska-low}\citep[SKA-LOW,][]{santos2015}.  These surveys allow for statistical detection of the 21 cm brightness fluctuations as a function of scale, which provides insights into the physical conditions in the IGM, the characteristic sizes of the HII regions [``bubbles'', \citet{furlanetto2006}] that grow and merge during reionization,  and the nature of the  luminous sources (quasars and galaxies) that contributed to the process.

In addition to 21 cm,  mapping the atomic and molecular emission line intensities  \citep[for reviews, see, e.g., ][]{bernal2022, kovetz2019} in the microwave and submillimetre regime (GHz and THz frequencies) constitute a novel window into the period of reionization,  as they are sensitive to the evolution of the star-formation rate \citep[SFR; e.g.][]{delooze2014, schaerer2020, romano2022} and molecular gas density \citep[e.g.,][]{zanella2018, dessauges2020} at high  redshifts.  Prominent among these are  the redshifted fine structure line of singly ionized carbon, [CII] with a rest wavelength $157.7 \mu$m \citep{silva2015, LidzTaylor16, fonscea2017, yue2015, dumitru2019, sun2018, gong2012, serra2016, li2015, 
crites2017, dizgah2019, hpcii2019},  and the 88 $\mu$m line of doubly ionized oxygen, [OIII] \citep{yang2020, jones2020, hpoiii}.
Various upcoming surveys, such as the the balloon-borne EXperiment for Cryogenic Large-Aperture Intensity Mapping 
\citep[EXCLAIM;][]{pullen2022}, the  CarbON CII line in 
post-rEionization and ReionizaTiOn \citep[CONCERTO;][]{serra2016, lagache2018}\footnote{https://people.lam.fr/lagache.guilaine/CONCERTO.html} , the Tomographic Intensity Mapping Experiment \citep[TIME; 
e.g.,][]{crites2014, crites2017}, and the Fred Young Submillimetre Telescope \citep[FYST;][]{terry2019}\footnote{https://www.ccatobservatory.org/index.cfm} aim to map the evolution of [CII] and [OIII] intensity over $ z \sim 3-8$. In \citet{hpcii2019}, the Herschel PACS observations of the 
Luminous Infrared Galaxies in the Great Observatories All-sky LIRG Survey 
\citep{hemmati2017} were combined with the intensity mapping 
measurement using Planck  High Frequency Instrument maps cross-correlated with BOSS 
quasars and CMASS galaxies from SDSS-III at $z \sim 2.6$ \citep{pullen2018}, to derive constraints on  the abundance and clustering of [CII] over $z \sim 0-4$.  
Both the [CII] and [OIII] line transitions have now been detected in galaxies out to $z \sim 9$, chiefly with the Atacama Large Millimetre/submillimetre Array \citep[ALMA;][]{carniani2017, laporte2017, harikane2020, tamura2019, hashimoto2018, katz2017, hashimoto2018a, walter2018,inoue2016}. The sensitivity of the ALMA has allowed,  for the first time, a characterization of the [CII] luminosity functions over  $z \sim 4.5 - 6$ from the ALMA Large Program to INvestigate CII at Early times   \citep[ALPINE;][]{yan2020, loiacono2021} survey and  at $z \sim 7$  from the Reionization Era Bright Emission Line Survey  \citep[REBELS;][]{oesch2022, bouwens2022} along with its pilot programs \citep{smit2018, schouws2022}.

Cross-correlating the high-redshift 21 cm survey with that of a different tracer (such as galaxies or atomic/molecular line maps) promises several advantages to both measurements \citep[e.g.,][]{visbal2010, carilli2011}.   Galaxy surveys cross-correlated with 21 cm intensity maps have the potential to break degeneracies in the constraints on the inferred astrophysical parameters \citep{dumitru2019} and scenarios \citep{wyithe2007} of reionization.  Moreover,  the foregrounds,  which currently are challenging to remove or avoid in 21 cm data analysis \citep[for a review, see, e.g.,][]{liu2020} largely originate from low redshifts, so the  probability of having shared foregrounds between reionization-era surveys is extremely low.  Thus, cross-correlation with known tracers at high redshifts can result in an unambiguous confirmation of the 21 cm detection \citep[][]{furlanetto2007},  with an improved signal-to-noise due to the non-correlation of systematics between the surveys. In the case of cross-correlation  with a  photometric galaxy survey smaller than a few square degrees,  information can be lost in the areas of the cross-power spectrum most affected by foregrounds \citep[e.g.,][]{lidz2009}. This is largely mitigated by using intensity mapping surveys with other line  tracers, such as [CII], covering a few ten square degrees or more \citep[e.g.,][]{beane2018, beane2019}. 

In this paper, we examine to what extent the current observational constraints on [CII] and [OIII] in galaxies during the epoch of reionization impact the sensitivities of a cross-correlation measurement between 21 cm surveys and the sub-mm regime. { While such cross-correlations have been analysed in some detail for observations of the post-reionization Universe ($z < 5$) in the sub-mm regime using facilities like SPHEREx \citep{gong2017}, we follow up on earlier work by \citet{wyithe2007, lidz2008, gong2012, furlanetto2007, kulkarni2016, dumitru2019} that laid out the theoretical basis and associated challenges at higher redshifts.}  We work with three latest compilations of the luminosity function from surveys targeting high-$z$ [CII] in individual galaxies: the ALPINE serendipitous and targeted detections over $z \sim 4.5 - 6$ \citep{yan2020, loiacono2021, oesch2022} and the REBELS galaxies at $z \sim 7$ from \citet{oesch2022}.  In so doing, we extend the parametrization developed in \citet{hpcii2019} to model the [CII] luminosity - halo mass relation over $z \sim 4.5-7$, finding  it to be well-described by a double-power law, with the best-fitting parameters in agreement with the results of earlier observations and simulations.  For the experimental configurations, we  use the improved versions of the FYST-like and EXCLAIM-like configurations introduced in \citet{hpoiii}, cross-correlated with a  21 cm survey from the Murchison Widefield Array (MWA), and its successor, the Square Kilometre Array (SKA)-LOW. We find that the new experiments should be able to make secure detections of the [CII] autocorrelation power over $z \sim 5-7$. We also forecast the detection of 21 cm - [CII] and 21 cm - [OIII] cross-power at $z \sim 5,6$ and 7,  finding that  the cross-correlation leads to significant improvements in the signal-to-noise ratio, which goes up to a few tens for the MWA-FYST-like experiment combination and to a few for the MWA-EXCLAIM-like survey. 

The paper is organized as follows. In Sec. \ref{sec:formalism},  we describe the formalism for modelling the occupation of HI in dark matter haloes which is used to derive the 21 cm autocorrelation power spectrum. We summarize the recent ALMA observations used for constraining the [CII] luminosity - dark matter halo mass relation over $z \sim 4.5 - 7$ using the abundance matching procedure, the resultant model parameters 
and their associated uncertainties. In Sec. \ref{sec:autocorr}, we derive the 21 cm and [CII]/[OIII] autocorrelation power spectra  and illustrate their observability with upcoming experiments. Finally, we describe (in Sec. \ref{sec:crosscorr}) the cross-correlation of the 21 cm surveys with the MWA and SKA-LOW with the sub-millimetre experiments probing $z \sim 4.5-7$, and forecast their expected sensitivities. We summarize our results and discuss the outlook in Sec. \ref{sec:conclusions}. We adopt the  $\Lambda$CDM 
cosmology with the parameters $h = 0.71, \Omega_b = 0.046, \Omega_m 
= 0.281, \sigma_8 = 0.8, \Omega_{\Lambda} = 0.719, n_s = 0.963$ throughout the paper.

\begin{table}
\begin{tabular}{lll}
{\bf \large Summary of model parameters}  & & \\
&&\\
 \noindent\rule{\linewidth}{0.1pt}
 & & \\
  {\bf \large HI Mass - halo mass}          &  & \\
  {\underline{Fitting function:}} &  &  \\
  & &  \\        
      $M_{\rm HI} (M) = \alpha f_{\rm H,c} M \left(M/10^{11} h^{-1} M_{\odot}\right)^{\beta} \exp\left[-\left(v_{c,0}/v_c(M)\right)^3\right]$;   & &   \\ 
 $\rho_{\rm HI} (r) =\rho_0 \exp(-r/r_s)$; & &  \\
 $c_{\rm HI}(M,z) \equiv \displaystyle{\frac{R_v (M)}{r_s}} = c_{\rm HI, 0} \displaystyle{\left(\frac{M}{10^{11} M_{\odot}}\right)^{-0.109}} 4/(1+z)^{\gamma}$ &  &  \\ 
  $v_c(M) = \displaystyle{\left(\frac{G M}{R_v(M)}\right)^{1/2}}$; $R_v(M)$ follows \eq{virialradius}, & & \\
  & & \\
  $f_{\rm H,c}$ follows \eq{fhc}. & & \\
&&\\
{\underline{Parameter values:}} & & \\
&&\\
$c_{\rm HI,0} = 28.65 \pm 1.76$   &  &                                                           
\\
$\alpha = 0.09 \pm 0.01$  &  &  
\\
log$_{10} (v_{c,0}/\rm km \ s^{-1}) = 1.56 \pm 0.04$      &  &                                                           
\\
$\beta = -0.58 \pm 0.06$       &  &                                                                
\\
$\gamma = 1.45 \pm  0.04$      &  &  
\\
 \noindent\rule{\linewidth}{0.1pt}
 & & \\
     {\large {\bf [CII] Luminosity - Halo mass} } &   &  \\
     \\
      {\underline{Fitting functions (all masses in $M_{\odot}$, luminosities in $L_{\odot}$):}} &  & \\
     && \\
     ${z \sim 0-3}$, fit to \citet{hemmati2017, pullen2018} & & \\
     $L_{\rm CII}(M,z) = \displaystyle{\left(\frac{M}{M_1}\right)}^{\beta} \exp(-N_1/M)\left(\frac{(1+z)^{2.7}}{1 + [(1+z)/2.9)]^{5.6}} \right)^{\alpha} $ & & \\ 
  {\underline{Parameter values:}} &  & \\
$M_1 = (2.39 \pm 1.86) \times 10^{-5}$ &&\\
$N_1 = (4.19 \pm 3.27) \times 10^{11}$; && \\
 $\beta = 0.49 \pm 0.38$ &&\\
  $\alpha = 1.79 \pm 0.30$ &&\\
 &&\\
  $z \sim 4.5-6$  ALPINE serendipitous, \citet{loiacono2021} & & \\
     $L_{\rm CII}(M,z) = 2N_1 M [(M/M_1)^{-\beta} + (M/M_1)^{\gamma}]^{-1} $ & & \\ 
  {\underline{Parameter values:}} &  & \\
$M_1 = (4.42 \pm 0.29) \times 10^{11}$; $N_1 = (1.92\pm 0.13) \times 10^{-2}$;
&& \\
 $\beta = 1.15 \pm 0.08$ ; $\gamma = 0.57 \pm 0.04$ &&\\
  &&\\
  $z \sim 4.5-6$  ALPINE targeted, \citet{oesch2022} & & \\
     $L_{\rm CII}(M,z) = 2N_1 M [(M/M_1)^{-\beta} + (M/M_1)^{\gamma}]^{-1} $ & & \\ 
  {\underline{Parameter values:}} &  & \\
$M_1 = (1.26 \pm 0.16) \times 10^{12}$; $N_1 = (8.04\pm 0.99) \times 10^{-4}$;
&& \\
 $\beta = 1.08 \pm 0.13$ ; $\gamma = 0.49 \pm 0.06$ &&\\
  &&\\
  $z \sim 7$ REBELS, \citet{oesch2022}  & & \\
      $L_{\rm CII}(M,z) = 2N_1 M [(M/M_1)^{-\beta} + (M/M_1)^{\gamma}]^{-1} $ & & \\ 
  {\underline{Parameter values:}} &  & \\
$M_1 = (9.79 \pm 0.09) \times 10^{12}$; $N_1 = (2.01\pm 0.19) \times 10^{-3}$;
&& \\
 $\beta = 0.94 \pm 0.09$ ; $\gamma = 0.51 \pm 0.05$ &&\\
\noindent\rule{\linewidth}{0.1pt}
 &&\\
    {\large {\bf [OIII] Luminosity - Halo mass}}  &   &  \\
   $z \sim 6-9$ & & \\
       {\underline{Fitting function:}} &  & \\
      &&\\
     $\log\displaystyle{\left(\frac{L_{\rm OIII}}{L_{\odot}}\right)} = 0.97 
\times \log \displaystyle{\frac{\rm SFR (M,z)}{[M_{\odot} \text{yr}^{-1}]}} + 7.4 $; &&\\
${\rm SFR}(M,z)$ follows \citet{behroozi2019}  \  & & \\ 
&&\\
\noindent\rule{\linewidth}{0.1pt}
 &&\\
\label{table:constraints}
\end{tabular}
\caption{Summary of the tracer-halo mass relations used in this work. For each tracer, we list the 
best-fitting functional form, parameter(s) constrained, and the redshift range of applicability.} 
\end{table}

\section{Relevant equations}
\label{sec:formalism}
We model the abundance and clustering of HI following the halo model framework developed in \citet{hparaa2017} and summarized briefly below. The average HI mass in a dark matter halo of mass $M$ at redshift $z$ is denoted by $M_{\rm HI}(M,z)$ and described by:
\begin{equation}
M_{\rm HI} (M, z) = \alpha f_{\rm H,c} M \left(M/10^{11} h^{-1} M_{\odot}\right)^{\beta} \exp\left[-\left(v_{c,0}/v_c(M)\right)^3\right]
\label{hihm}
\end{equation}
in which the circular velocity, $v_c (M)$ is defined in terms of the virial radius $R_v(M)$ of the halo, 
 \begin{equation}
    v_{c}(M) = \sqrt\frac{GM}{R_v(M)}
    \label{vcRv}
\end{equation}
and
\begin{equation}
 R_v (M) = 46.1 \ {\rm{kpc}} \  \left(\frac{\Delta_v \Omega_m h^2}{24.4} \right)^{-1/3} \left(\frac{1+z}{3.3} \right)^{-1} \left(\frac{M}{10^{11} M_{\odot}} \right)^{1/3}
 \label{virialradius}
 \end{equation} 
The cosmological hydrogen fraction is denoted by $f_{\rm H,c}$ and given by
\begin{equation}
f_{\rm H,c} = (1 - Y_{\rm He}) \Omega_b/\Omega_m \, ,
\label{fhc}
\end{equation}
in which $Y_{\rm He} = 0.24$ is the primordial helium fraction. The free parameters in \eq{hihm} are (a) $\alpha$, the fraction of HI (relative to cosmic) in the dark matter halo,  (b) $\beta$, the logarithmic slope of the relation and (c) $v_{\rm c,0}$,  which represents the minimum (circular) velocity required for a halo to host hydrogen, and is related to the strength of the UV field that prevents the efficient cooling of gas in low mass haloes \citep{rees1986, efstathiou1992, barnes2014}.

The distribution of the HI gas as a function of scale is described by a profile, whose functional form is given by:
\begin{equation}
\rho_{\rm HI}(r,M) = \rho_0 \exp(-r/r_s)
\label{rhodefexp}
\end{equation}
in which $r_s$ is the scale factor, which is related to the virial radius by the concentration parameter
$c_{\rm HI}(M,z)$ defined as \citep{maccio2007}:
\begin{equation}
 c_{\rm HI}(M,z) =  c_{\rm HI, 0} \left(\frac{M}{10^{11} M_{\odot}} \right)^{-0.109} \frac{4}{(1+z)^{\gamma}}.
\end{equation} 
with the constant $\rho_0$ in Eq. (\ref{rhodefexp}) fixed by normalizing the profile within $R_v (M)$ to be equal to $M_{\rm HI}$. 
 
The HI power spectrum is derived analogously to the dark matter case by using 
 the Fourier transform of the profile function:
\begin{equation}
 u_{\rm HI}(k|M) = \frac{4 \pi}{M_{\rm HI} (M)} \int_0^{R_v} \rho_{\rm HI}(r) \frac{\sin kr}{kr} r^2 \ dr
\end{equation}
from which the one- and two-halo terms can be obtained through:
\begin{equation}
P_{\rm 1h, HI} =  \frac{1}{\bar{\rho}_{\rm HI}^2} \int dM \  n(M) \ M_{\rm HI}^2 \ |u_{\rm HI} (k|M)|^2
\label{onehalo}
\end{equation}
and 
\begin{equation}
P_{\rm 2h, HI} =  P_{\rm lin} (k) \left[\frac{1}{\bar{\rho}_{\rm HI}} \int dM \  \frac{dn}{dM} \ M_{\rm HI} (M) \ b (M) \ |u_{\rm HI} (k|M)| \right]^2
\label{twohalo}
\end{equation}
where ${dn}/{dM}$ is the halo mass function [we use the \citet{sheth2002} form]. In the above expressions, $\bar{\rho}_{\rm HI}(z)$ is
the average HI density in the halo defined by:
\begin{equation}
\bar{\rho}_{\rm HI}(z)= \int_{M_{\rm min}}^{\infty} dM \frac{dn}{dM} M_{\rm HI}(M) \,
\end{equation}
The total HI power spectrum is then given by 
\begin{equation}
P_{\rm HI} = P_{\rm 1h, HI} + P_{\rm 2h, HI}.
\label{powerspechi}
\end{equation}

The free parameters in the framework, $\alpha, \beta, v_{c,0}, c_{\rm HI}$ and $\gamma$ are obtained by a detailed fit to the ensemble of HI data over $z \sim 0-5$, including galaxy surveys at $z \sim 0$ and the Damped Lyman Alpha observations at $z \sim 2-5$ \citep{hparaa2017}. The best fitting values and uncertainties are summarized in Table \ref{table:constraints}. We assume that the form of the HI-halo mass relation remains relatively unchanged over $z \sim 5-7$, noting that doing so neglects the influence of the bubbles/diffuse neutral regions in the IGM and the spatially varying strength of the ionizing radiation field,  both of which are expected to introduce scale dependence. The wide range of possibilities for these parameters depending on the reionization scenario, however [e.g., the scale of transition between the anti-correlation and correlation of the [CII]-HI signal varying by over an order of magnitude, as noted by \citet{dumitru2019}] makes their values practically unconstrained with current data. 
In the light of recent observations and simulations \citep[for a review, see, e.g., 
][]{peroux2020}, which suggest that the total HI abundance shows no or mild evolution to $z \sim 6-7$, the present approach may be a good approximation at least for the magnitude of the expected HI signal. We revisit this point in the context of theoretical expectations in Sec. \ref{sec:conclusions}.

The measured signal in a 21 cm intensity mapping experiment is the three dimensional power spectrum of the fluctuations,  given by \citep[e.g.,][]{battye2012}:
\begin{equation}
[\delta T_{\rm HI} (k,z)]^2  = \bar{T}(z)^2 P_{\rm HI}(k,z)
\end{equation}
in which the mean brightness temperature at redshift $z$ and frequency $\nu = 1420{\rm  MHz}/ (1+z)$ is given by\footnote{We assume that peculiar velocities are negligible, that the line profile is very narrow and that absorption can be  neglected, which is a good approximation in the mid-to end stages of reionization when the spin temperature of the gas is much higher than the CMB temperature \citep[e.g.,][]{pritchard2007, mcquinn2006}.}:
\begin{eqnarray}
 \bar T (z) &=& \frac{3 h_{\rm Pl} c^3 A_{10}}{32 \pi k_B m_p^2 \nu_{21}^2}\frac{(1+z)^2}{H(z)} \Omega_{\rm HI} (z) \rho_{c,0} \nonumber \\
            &\simeq& 44 \ \mu {\rm{K}} \left(\frac{ \Omega_{\rm HI} (z) h }{2.45 \times10^{-4}}\right)\frac{(1+z)^2}{E(z)},
  \label{tbar}
\end{eqnarray}
where $H(z)$ is the Hubble parameter at redshift $z$, $E(z) = H(z) / H_0$,  $\Omega_{\rm HI}(z)$ is the comoving density parameter of HI relative to the present-day critical density $\rho_{c,0}$,  $A_{10}$ is the Einstein-A coefficient for the spontaneous emission between the lower (0) and upper (1) levels of the hyperfine splitting of HI, $\nu_{21} = 1420$ MHz is the rest frequency  of the emission, and other symbols have their usual meanings.
As in previous literature \citep[e.g.,][]{lidz2008}, we work with the power spectrum per unit interval in logarithmic $k$-space,
\begin{equation}
\Delta_{\rm HI}^2 = \frac{k^3 [\delta T_{\rm HI} (k,z)]^2}{2 \pi^2} =  \bar T (z)^2  \frac{k^3 P_{\rm HI} (k,z)}{2 \pi^2} 
\end{equation}
which can be expressed in units of mK$^2$ or equivalently in terms of the intensity, $I_{\nu, {\rm HI}}$ [which is proportional to $\bar{T}(z)$ in the Rayleigh limit] in units of (Jy/sr):
\begin{equation}
I_{\nu,{\rm HI}} = \frac{2 k_B \bar{T}(z)}{\lambda^2} 
\label{tempintensityrayleigh}
\end{equation}
where $\lambda \equiv c/\nu = 21 {\rm cm}(1+z)$.

The abundance and clustering of the submillimetre tracers ([CII]/[OIII]) are also modelled in terms of their observed intensities [we follow the procedure described in \citet{hpoiii}]:  
\begin{equation}
 I_{\nu_{\rm obs}, {i}} = \frac{c}{4 \pi} \int_0^{\infty} dz' 
\frac{\epsilon[\nu_{\rm obs, i} (1 + z'), z']}{H(z') (1 + z')^4}
\end{equation} 
In the above expression, the emissivity, $\epsilon(\nu, z)$ of the tracer is given by:
\begin{equation}
 \epsilon(\nu, z) = \delta_{\rm D}(\nu - \nu_{i}) (1 + z)^3 \int_{M_{\rm min, 
i}}^{\infty} dM \frac{dn}{dM} L_{i}(M,z)
\label{epsilonciioiii}
\end{equation} 
In the above, the $L_{\rm i}$ represents the average luminosity of the line-emitting (with $i \in \{\rm [CII], [OIII]\}$) galaxies as a function of their halo mass $M$ and redshift $z$. The emitted and observed frequencies are denoted by $\nu_{i}$ and $\nu_{{\rm obs}, i}$ respectively,  with $M_{\rm min, i}$ being the minimum halo mass associated with [CII]/[OIII] line emission. 
Using \eq{epsilonciioiii}, the intensity of emission can be re-expressed as:
\begin{equation}
I_{\nu_{\rm obs}, i} = \frac{c}{4 \pi} \frac{1}{\nu_{i} H(z)}  
\int_{M_{\rm min, i}}^{\infty} dM \frac{dn}{dM} L_{i}(M,z)
\label{CIIspint}
\end{equation} 
{ The value of $M_{\rm min}$ stands for the minimum mass of the haloes that can form stars and produce line emission, which is close to the atomic hydrogen cooling limit, situated between $10^8$ and $10^9 M_{\odot}$ at these redshifts. Throughout this work, we use the fiducial value of $M_{\rm min} = 10^9 h^{-1} M_{\odot}$ for both [OIII] and [CII] \citep[following previous work in][]{munoz2011, chung2020, hpoiii}. A lower value causes a change in $P(k)$ by factors of order a few \citep{chung2020}.}

The measured power spectrum of the intensity is composed of (i) a shot noise term, defined through:
\begin{equation}
 P_{\rm shot, i}(z) = \frac{\int_{M_{\rm min,i}}^{\infty} dM (dn/dM) L_{i} 
(M,z)^2}{\left(\int_{M_{\rm min, i}}^{\infty} dM (dn/dM) L_{i} 
(M,z)\right)^2}
\end{equation}
and (ii) a clustering term, which contains the large-scale bias:
\begin{equation}
 b_{i}(z) = \frac{\int_{M_{\rm min, i}}^{\infty} dM (dn/dM) L_{i} 
(M,z) b(M,z)}{\int_{M_{\rm min, i}}^{\infty} dM (dn/dM) L_{i} (M,z)}
\end{equation} 
of the tracer with respect to the underlying dark matter. In the above, $b(M,z)$ denotes the dark 
matter halo bias following \citet{scoccimarro2001}. The full power spectrum of the line intensity is given by the sum of the clustering and shot noise components:
\begin{equation}
 P_{i}(k,z) =  I_{\nu_{\rm obs}, i} (z)^2 [b_{i}(z)^2 P_{\rm lin}(k,z) + 
P_{\rm shot, i}(z)]
\label{powerspeccii}
\end{equation} 
in which  $P_{\rm lin}(k,z)$ is the linear matter power spectrum.

The luminosity to host halo mass relation, $ L_{i} 
(M,z)$ is thus the primary astrophysical ingredient used to derive the power spectrum for submillimetre tracers.
In  \citet{hpcii2019},  this quantity was constructed by using  the luminosity function at $z \sim 0$ from the Herschel PACS observations of the 
Luminous Infrared Galaxies in the Great Observatories All-sky LIRG Survey   \citep{hemmati2017}, and the intensity mapping cross-correlation measurement at $z \sim 2.6$ using the Planck maps combined with BOSS QSOs and CMASS galaxies.  The average [CII] luminosity - halo mass relation at $z \sim 0$ was derived by matching the measured luminosity function, $\phi(L_{\rm CII})$ to the abundance of dark matter haloes [see also \citet{hpgk2017, hpco, chung2022} for examples of using this procedure for other emission line tracers]:
\begin{equation}
  \int_{M (L_{\rm CII})}^{\infty} \frac{dn}{ d \log_{10} M'} \ d \log_{10} M' = \int_{L_{\rm CII}}^{\infty} \phi(L_{\rm CII}) \ d \log_{10} L_{\rm CII}
  \label{abmatchcii}
\end{equation}
in which $dn / d \log_{10} M$ is the number density of dark matter haloes having logarithmic
masses between $\log_{10} M$ and $\log_{10} (M$ + $dM)$, and $\phi(L_{\rm CII})$ is the
corresponding number density of [CII]-emitting galaxies in logarithmic luminosity bins.  Solving
Equation~(\ref{abmatchcii}) gives a relation between the [CII] luminosity
$L_{\rm CII}$ and the halo mass $M$, assuming it to be monotonic (which is a reasonable assumption in the light of the current data).
The best-fitting functional form obtained at $z \sim 0$ was then applied to the $z \sim 3$ intensity mapping measurement to constrain its evolution:
\begin{equation}
L_{\rm CII}(M,z) = \left(\frac{M}{M_1}\right)^{\beta} \exp(-N_1/M) 
\left(\frac{(1+z)^{2.7}}{1 + [(1+z)/2.9)]^{5.6}} \right)^{\alpha}
\label{lciihighz}
\end{equation}
over $z \sim 0-3$. The values and uncertainties on the free parameters $\alpha, \beta, M_1$ and $N_1$ are summarized in Table \ref{table:constraints}.

\begin{figure}
    \centering
    \includegraphics[width = \columnwidth]{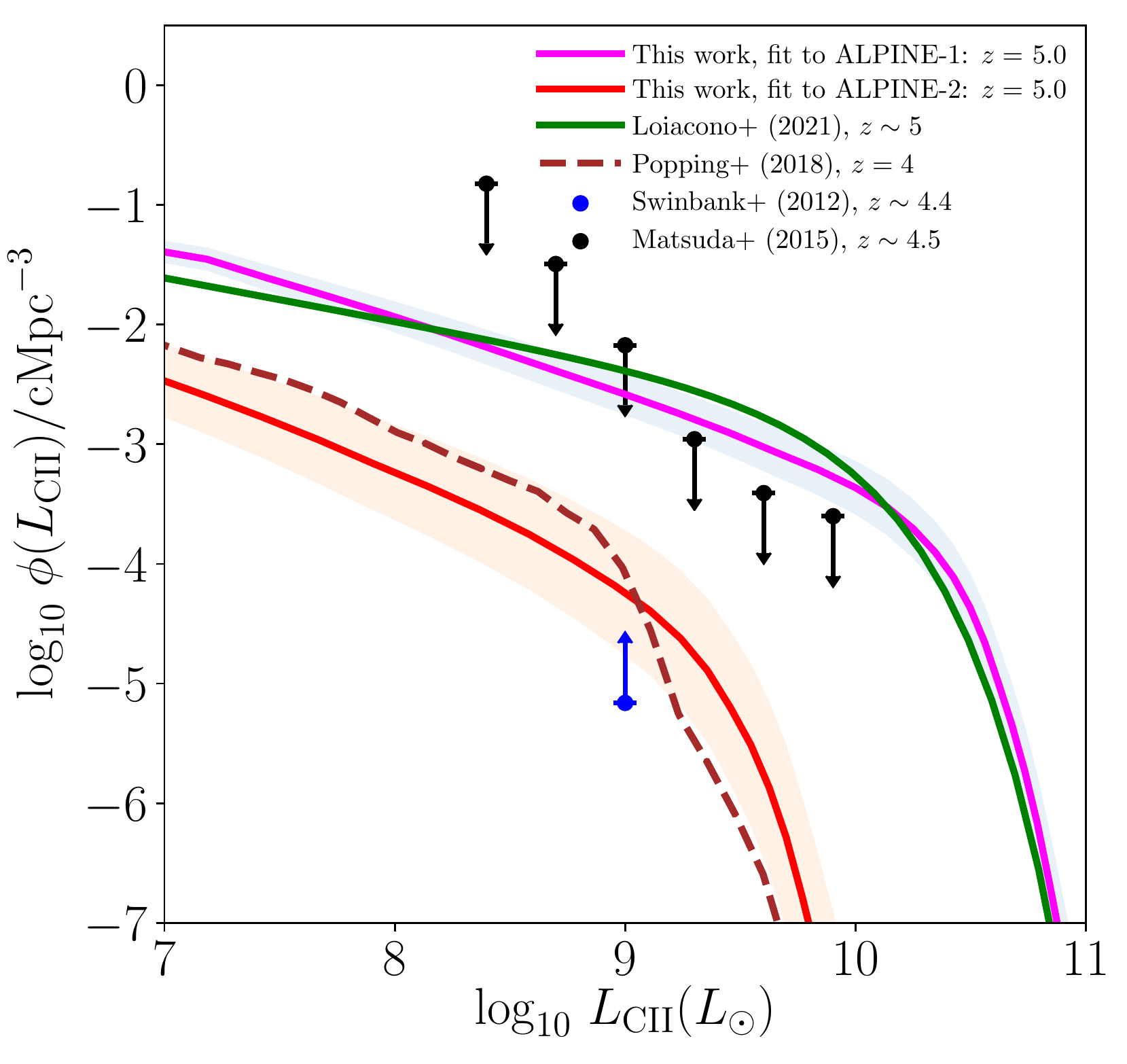}  \includegraphics[width = \columnwidth]{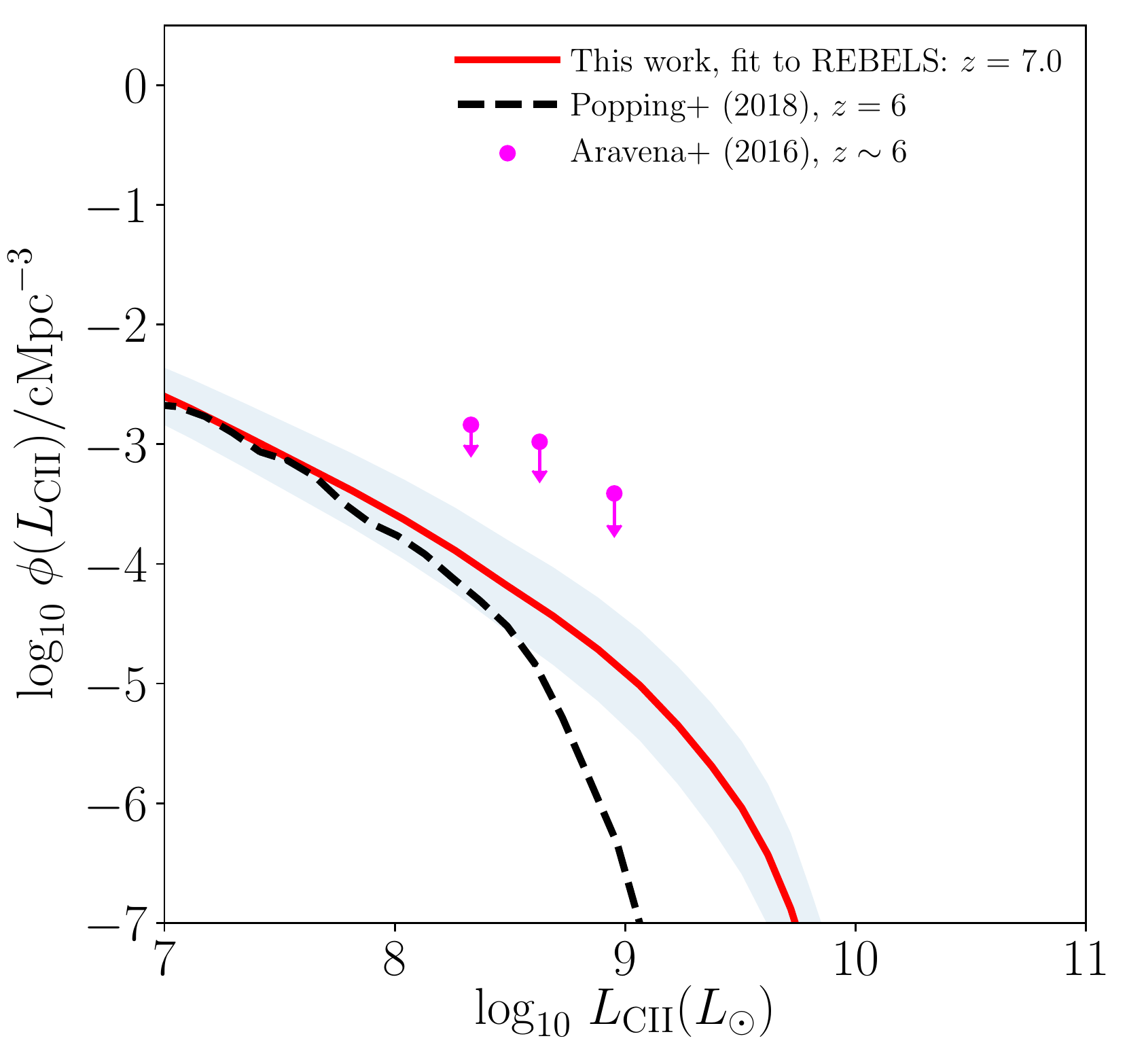}
    \caption{[CII] luminosity function derived from fitting the ALPINE clustered serendipitous and targeted detections, denoted by ALPINE-1 and ALPINE-2 respectively at $z \sim 4.5 - 6$ \citep[top panel;][]{loiacono2021, oesch2022} and the REBELS targets at $z \sim 7$ \citep[lower panel;][]{oesch2022} respectively.  The shaded regions indicate the range of vlaues obtained from the parameter uncertainties. Also shown are the measured luminosity function of clustered serendipitous ALPINE emitters at $z \sim 5$ \citep{loiacono2021},  the lower limits on the cumulative [CII]  luminosity function of two ALMA candidates at $z \sim 4.4$ \citep{swinbank2012}, the semi-analytic 
modelling results  of \citet{popping2019} at $z \sim 4$,  and the upper limits on the cumulative [CII] luminosity 
function at $z \sim 4.5$ based on (i) a blind search using ALMA archival data \citep{matsuda2015} and (ii)
the ASPECS blind survey candidate [CII] emitters at $z \sim 6-8$ \citep{aravena2016}.}
    \label{fig:ciilumfunc}
\end{figure}

We now extend the above formalism to derive the [CII] luminosity - halo mass relation at $z \sim 4.5 - 7$ by using the new targeted and serendipitous detections from the ALPINE \citep{yan2020, loiacono2021, oesch2022} survey at $z \sim 4.5 - 6$, and the  REBELS \citep{oesch2022} results for $z \sim 7$.

The ALMA Large Program to INvestigate CII at Early Times \citep[ALPINE;][]{lefevre2020, faisst2020, bethermin2020} survey  measured [CII] line emission in galaxies over $z \sim 4.5 - 6$ in the Extended Chandra Deep Field South and Cosmic Evolution Survey (COSMOS) fields. The recently explored ability of the ALMA to identify serendipitously detected galaxies in blind surveys allowed the characterization of the serendipitous line \citep{loiacono2021} and continuum \citep{gruppioni2020}  [CII] luminosity functions from ALPINE. { Galaxies in the serendipitous sample are especially luminous, with 11 out of 12 emitters in its `clustered' subsample belonging to a known overdensity in the COSMOS field. In contrast, the [CII] luminosity function  of targeted detections in the ALPINE field over $z \sim 4.5 - 6$ has also been calibrated using a different approach using the UV luminosity function in \citet{oesch2022}. These findings indicate that together, the clustered serendipitous and the targeted LFs can be considered upper and lower limits,  respectively, on the [CII] luminosity over $z \sim 4.5 - 6$ \citep[more details are presented in][]{oesch2022}.}

At $z \sim 7$, we use the ALMA  Reionization Era Bright Emission Line Search (REBELS) sample of UV-selected star forming galaxies at $z \sim 6.5-7.8$. The [CII] luminosity function at $z \sim 7$ from the REBELS survey may be considered a good representation of the $z \sim 7$ population as a whole \citep{oesch2022}, and is also found to agree well with the results of simulations.

Both the ALPINE and REBELS luminosity functions are well-described by Schechter forms \citep{schechter1976, yan2020, loiacono2021, oesch2022}. The best-fitting [CII] luminosity to host halo mass relations are obtained for the targeted and clustered serendipitous\footnote{We use the lower limit of the measured faint-end slope for the luminosity function of serendipitous emitters, as is needed to ensure monotonicity in the relation.} [CII] detections at $z \sim 4.5-6$ and the REBELS sample at $z \sim 7$  by using the abundance matching procedure of \eq{abmatchcii},  and the resultant values and error bars are summarized in Table \ref{table:constraints}.  Similar to the evolution of the CO luminosity to host halo mass \citep[e.g.,][]{chung2022, hpco}  a double power law form for $L_{\rm CII} (M,z)$ is found to best fit the $z \sim 4.5-7$ data. The luminosity functions derived from the fitting forms are plotted in Fig. \ref{fig:ciilumfunc}  at $z \sim 4-6$ and $z \sim 7$, along with  other high-redshift upper limits and simulation constraints on [CII].

For [OIII], we use the best-fitting functional form employed by \citet{hpoiii} and derived from the observations of the ALMA [OIII] detections in \citet{harikane2020}:
\begin{equation}
\log\left(\frac{L_{\rm OIII}}{L_{\odot}}\right) = 0.97 
\times \log \frac{\rm SFR}{[M_{\odot} \text{yr}^{-1}]} + 7.4
\label{oiiilumfunc}
\end{equation}
which connects the [OIII] luminosity to the star formation rate  (SFR) at $z \sim 6-9$. The empirical SFR - halo mass relation, ${\rm{SFR}}(M,z)$ from \citet{behroozi2019} is used to infer the $L_{\rm OIII}(M,z)$ using \eq{oiiilumfunc}.

\section{Autocorrelation forecasts for [CII]/[OIII] and HI}
\label{sec:autocorr}
\subsection{Sub-millimetre measurements}
\label{sec:autocorrsubmm}
We now forecast the expected auto-correlation signal and noise power spectra of HI and the sub-millimetre lines.For the experiments, we consider plausible improvements to the Fred Young Submillimetre Telescope (FYST)\footnote{https://www.ccatobservatory.org/index.cfm} located at the Cerro Chajnantor in Chile, covering the frequency range 212 - 428 GHz,  and the EXCLAIM \citep[EXperiment for Cryogenic Large-Aperture Intensity Mapping;][]{exclaimpaper2020, pullen2022}, a  balloon-based spectrometer mission surveying a $\sim 400$ sq. deg area overlapping with the Baryon Oscillation Spectroscopic Survey (BOSS), over the frequency range 420-540 GHz.  We consider both a Stage II concept as well as a more advanced Stage III/IV configuration for the improved FYST, following the specifications outlined in \citet{hpoiii}.
For the improved EXCLAIM-like experiment, we use the configuration considered in \citet{hpoiii} with an increased timescale of observation compared to the nominal survey. The parameters of the configurations considered are listed in Table \ref{table:improved}.

\begin{table*}
\begin{center}
    \begin{tabular}{ | c | c | c | c | c | c | c | c | c |  p{5 cm} |}
    \hline
    Configuration &  $D_{\rm dish}$ (m.) & $\Delta \nu$ (MHz) &  $N_{\rm spec, 
eff} $ & $S_{\rm A}$ (sq. deg.) & $\sigma_{\rm N}$ 
(Jy s$^{1/2}$ / sr) & $B_{\nu}$ & $t_{\rm obs}$ (h.)   \\ \hline

   FYST-like (Stage II)  &  9 & 400 &  1 & 100 & $4.84 \times 10^4$  & 40 GHz & 2000 h \\ 
   FYST-like (Stage III/IV)  &  9 & 400 &  16000 & 100 & $2.1 \times 10^5$  & 40 GHz & 2000 h  \\
       EXCLAIM-like & 0.74 & 1000 & 30 & 100 & $3 \times 10^5$  & 40 GHz  & 4000 h \\
\hline
    \end{tabular}
\end{center}
\caption{ {Experimental parameters for improved EXCLAIM-like and FYST-like survey designs, motivated from the analyses of \citet{hpoiii}, with an increased timescale for the EXCLAIM-like experiment.}}
 \label{table:improved}
\end{table*}

The improved FYST-like and EXCLAIM-like experiments are sensitive to [CII] and [OIII] respectively in the range $z \sim 5-7$.  
We follow \citet{hpoiii} in defining the noise in an autocorrelation survey in the sub-millimetre regime with the FYST-like or EXCLAIM-like configurations, given by:
\begin{equation}
P_{\rm N, ij} = V_{\rm pix, ij} \frac{\sigma_{\rm N, j}^2}{t_{\rm pix, ij}}
\label{noiseciiauto}
\end{equation}
in which the parameter $\sigma_{\rm N, j}$ is the noise-equivalent intensity (NEI) of the experiment $j \in \{\rm EXCLAIM, \ FYST\}$ as defined in Table \ref{table:improved}, and $i \in \{\rm [CII], [OIII]\}$. 
In the above,  $t_{\rm pix, ij}$ is given by
\begin{equation}
t_{\rm pix, ij} = t_{\rm obs, j} N_{\rm spec, eff, j} \frac{\Omega_{\rm beam, ij}}{S_{A, j}}
\label{eqntpixauto}
\end{equation}
with $t_{\rm obs, j}$ the total observation time of experiment $j$, and $N_{\rm spec, eff, j}$ the effective number of pixels of the experimental configuration \citep[e.g.,][]{hpcii2019}.  The $S_{A, j}$ is the survey area on the sky, the field of view is $\Omega_{\rm beam, ij} = 2 \pi (\sigma_{\rm beam, ij})^2 = 2 \pi (\theta_{\rm beam, ij}/\sqrt{8 \ln 2})^2$, with $\theta_{\rm beam, ij} = \lambda_i (1+z)/D_{\rm dish, j}$ being the full width at half maximum and $\lambda_i$  the rest wavelength of the respective transition $(i \in \rm [CII], [OIII])$. 
Also, the pixel volume, $V_{\rm pix, ij}$ is given by
\citep[e.g.,][]{dumitru2019}:
\begin{eqnarray}
V_{\rm pix, ij} &=& 1.1 \times 10^3  {\rm (cMpc}/h)^3 \left (\frac{\lambda_i}{158  \ 
\mu m} \right) \left(\frac{1 +z}{8} \right)^{1/2}  \nonumber \\
&& \left(\frac{\theta_{\rm beam, ij}}{10 '} \right)^2 \left(\frac{\Delta \nu_j}{400 
{\rm MHz}} \right)
\label{eqnvpix}
\end{eqnarray}
in which $\Delta \nu_j$ is the spectral resolution of the instrument. 

\begin{figure}
    \centering
    \includegraphics[width = 0.9\columnwidth]{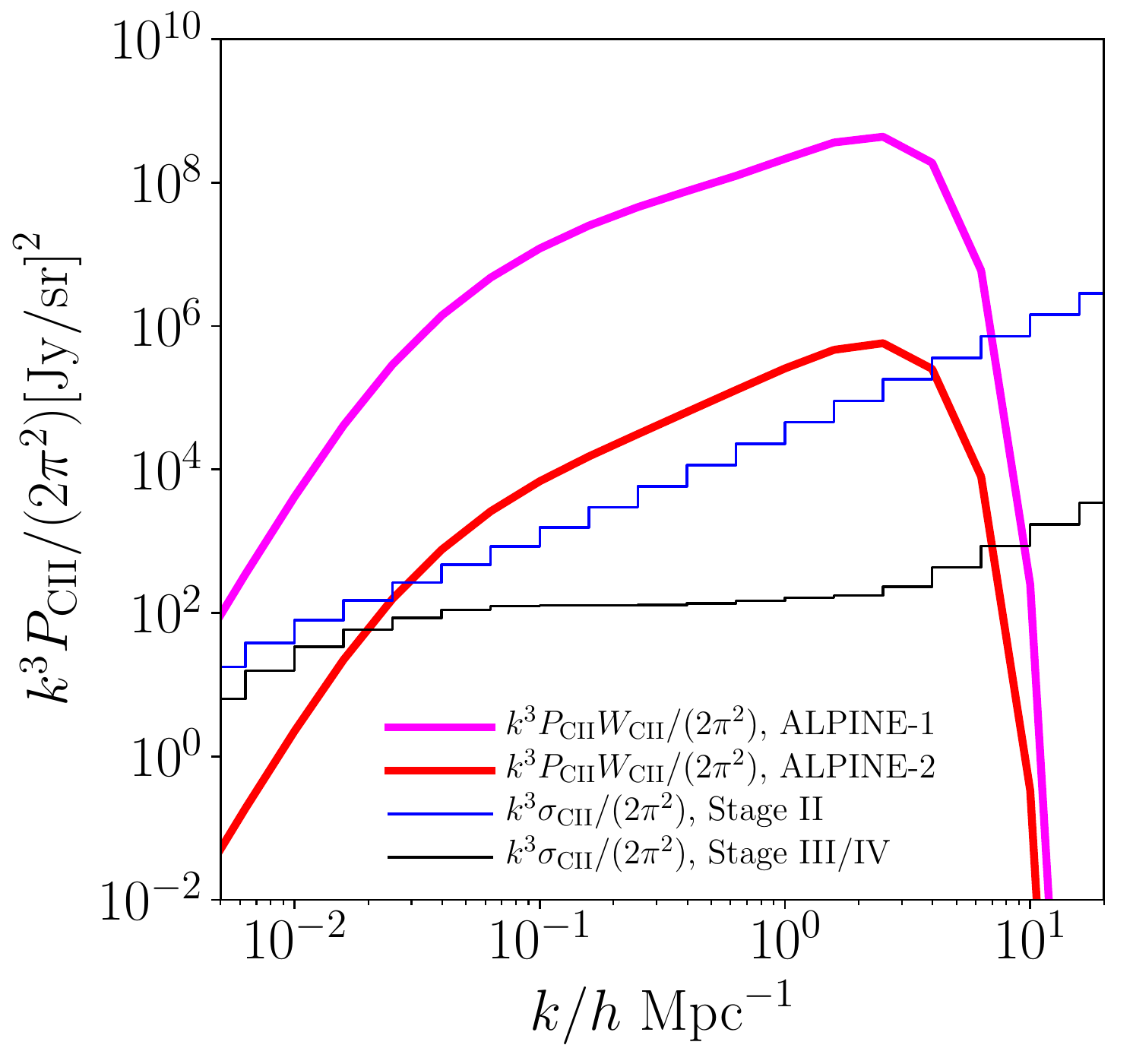} \includegraphics[width = 0.9\columnwidth]{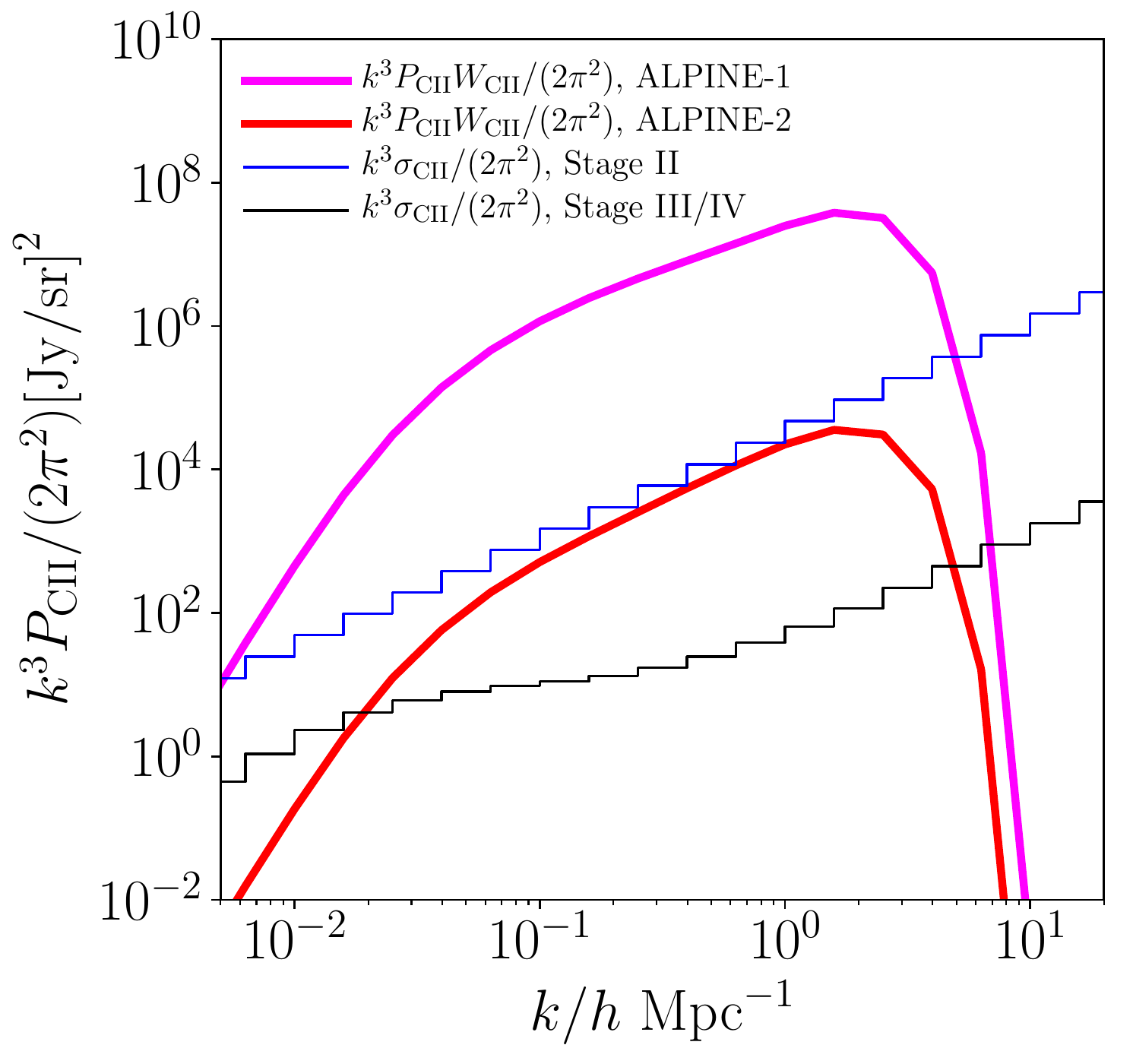}\\
    \includegraphics[width = 0.9\columnwidth]{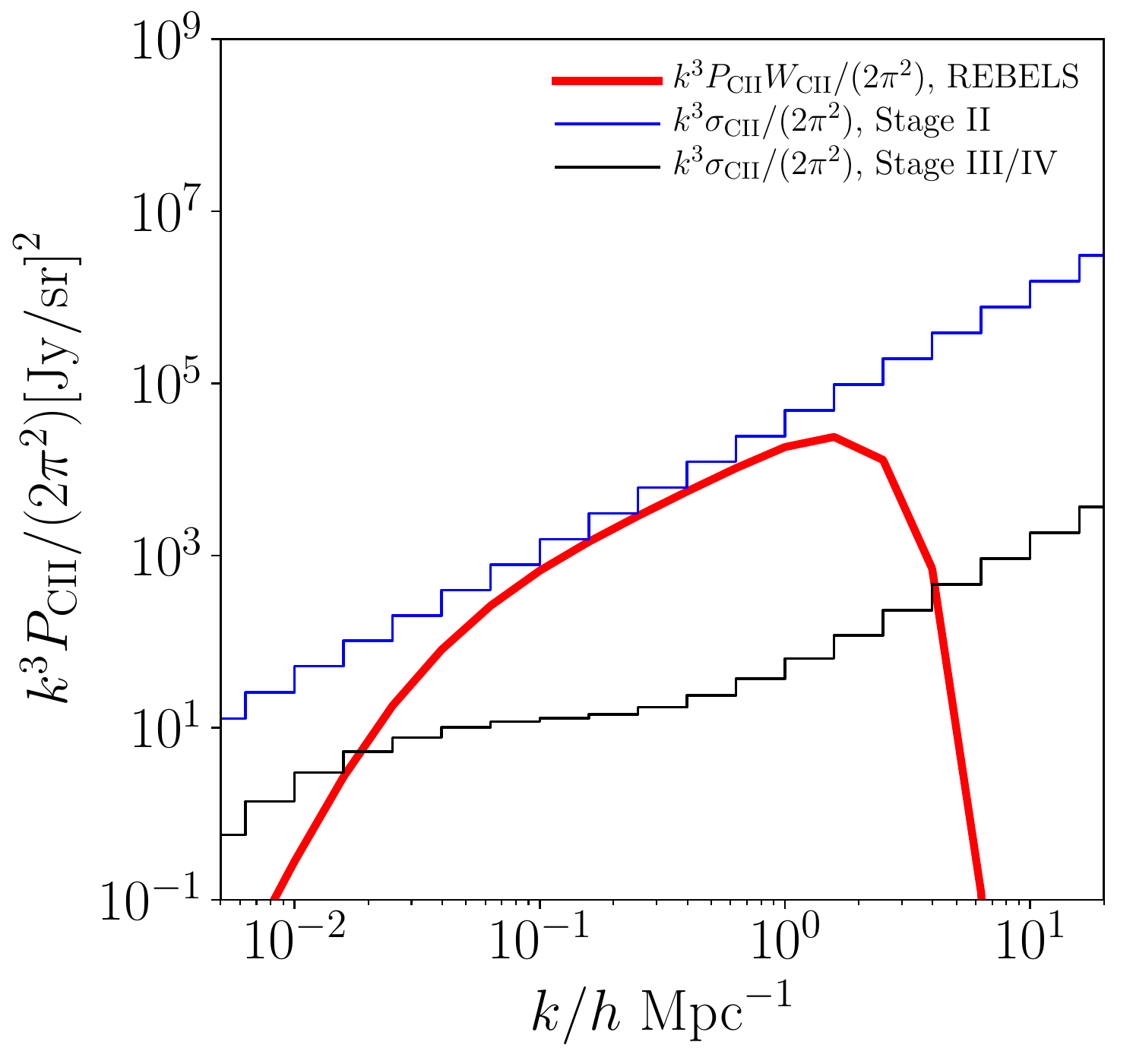}
    \caption{Auto-correlation signal  $k^3 P_{\rm CII}/(2 \pi^2)$ and expected noise $k^3 \sigma_{\rm CII}/(2 \pi^2)$, for improved versions of FYST-like Stage II and Stage III/IV surveys defined in Table \ref{table:improved}, probing [CII] in bands centred on $z \sim 5 , 6$ and 7. At $ z \sim 5$ and $z \sim 6$, the signal power is plotted for both the [CII]  luminosity - halo mass fits in Table \ref{table:constraints},  from the ALPINE serendipitous and targeted detections (denoted as ALPINE-1 and ALPINE-2 respectively). The [CII] power at $z \sim 7$ is derived from the fit to the REBELS targets.}
    \label{fig:autocorrcii}
\end{figure}

\begin{figure}
    \centering
    \includegraphics[width = \columnwidth]{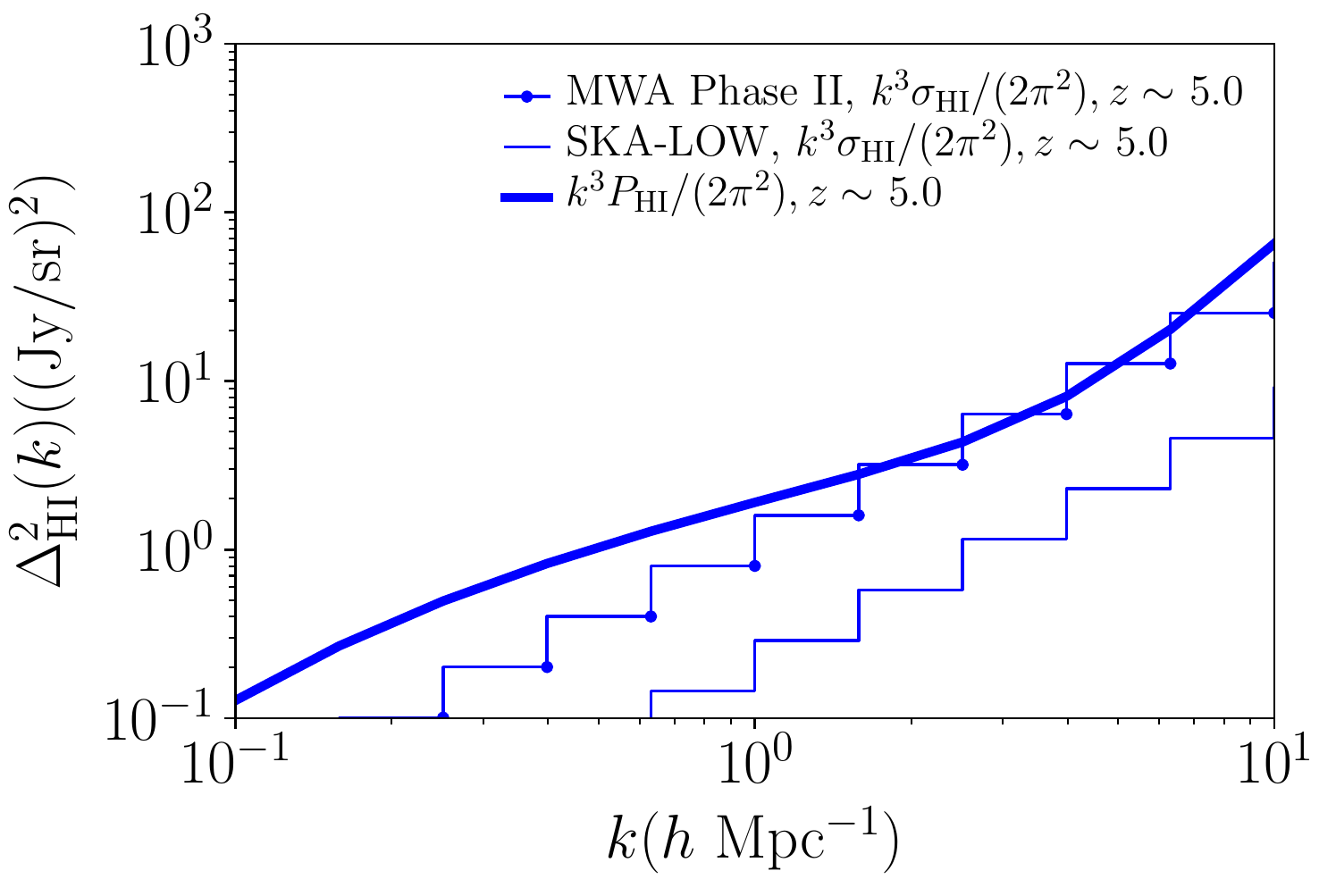}  \includegraphics[width = \columnwidth]{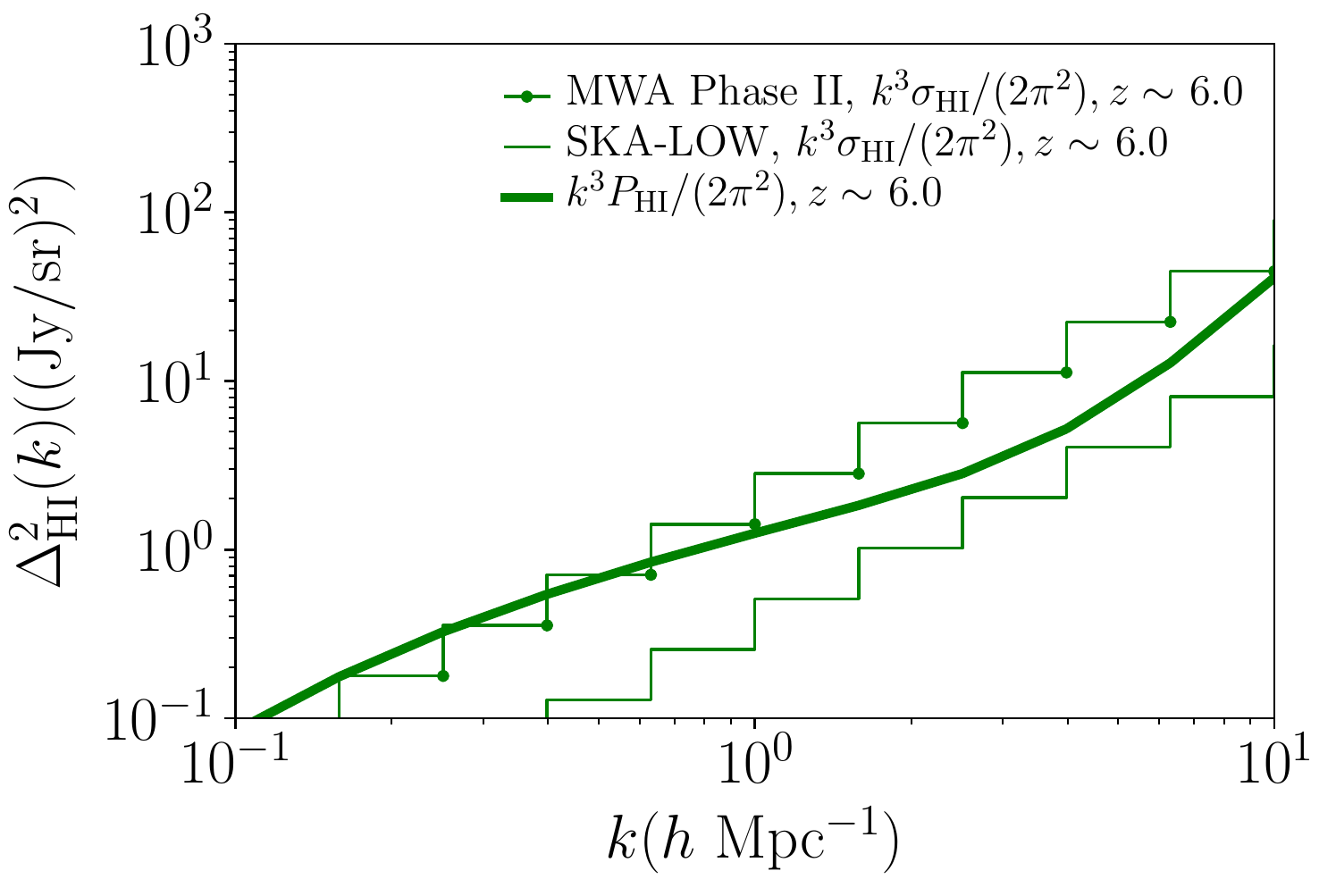} \includegraphics[width = \columnwidth]{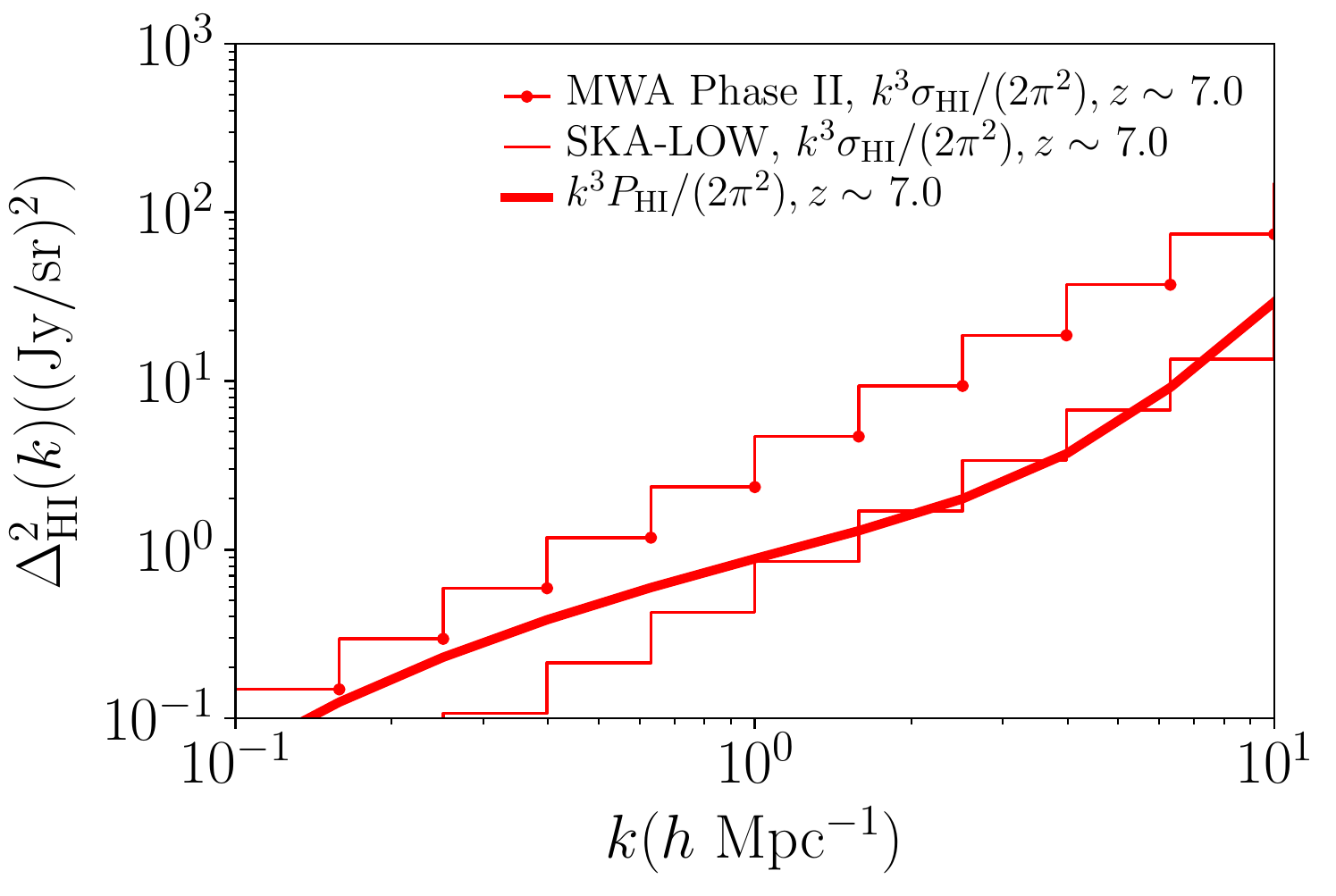}
    \caption{The 21 cm auto-correlation signal $k^3 P_{\rm HI}/(2 \pi^2)$ at $z \sim 5, 6, 7$ and expected  noise  $k^3 \sigma_{\rm HI}/(2 \pi^2)$  for the MWA (dotted step) and SKA-LOW (solid step) configurations described in Table \ref{table:hi}.}
    \label{fig:autocorrhi}
\end{figure}

\subsection{Finite spatial and spectral resolution}
\label{sec:volres}
We also correct for the effects due to finite spatial and spectral resolution of the beam in the  [CII] and [OIII] surveys.  Following  \citet{hpoiii}, the relevant window functions are defined by:
\begin{eqnarray}
 W_{\rm beam}(k) &=& e^{-k^2 \sigma_{\perp, j}^2} \int_0^1 e^{-k^2 \mu^2 (\sigma_{\parallel, ij}^2 - \sigma_{\perp, j}^2)} d \mu \nonumber \\
   \noindent &=& \frac{1}{k \sqrt{\sigma_{\parallel, ij}^2 - \sigma_{\perp, j}^2}} \frac{\sqrt{\pi}}{2} {\rm{Erf}} \left(k  \sqrt{\sigma_{\parallel, ij}^2 - \sigma_{\perp, j}^2} \right)  \nonumber \\
   &\times& \exp(-k^2 \sigma_{\perp, j}^2)
   \label{wij}
\end{eqnarray}
in which $\sigma_{\perp, j}$ and $\sigma_{\parallel, ij}$ are defined through \citep{li2015}:
\begin{equation}
    \sigma_{\perp, j} = \chi(z) \sigma_{\rm beam, j}
    \label{sigmaperp}
\end{equation}
where $\chi(z)$ is the comoving distance to redshift $z$, $\sigma_{\rm beam, j} = \theta_{\rm beam, ij}/\sqrt{8 \ln 2}$, and
\begin{equation}
    \sigma_{\parallel, ij} = \frac{c}{H(z)} \frac{(1+z)^2 \Delta \nu_j}{\nu_{\rm obs, i}}
    \label{sigmapar}
\end{equation}
The finite volume of the survey also introduces a correction factor, encapsulated by:
\begin{equation}
    \begin{split}
W_{\rm vol}(k,\mu) = & \left(1-\exp\left\lbrace -\left(\frac{k}{k^{\rm min}_\perp}\right)^2\left(1-\mu^2 \right) \right\rbrace \right)\times \\
& \times \left(1-\exp\left\lbrace -\left(\frac{k}{k^{\rm min}_\parallel}\right)^2\mu^2 \right\rbrace \right).
\end{split}
\label{eq:Wk_vol}
\end{equation}
In the above,  $L_\perp$ and $L_\parallel$ are  the maximum transverse and radial length scales probed by the survey,  $k^{\rm min}_{\perp} \equiv 2\pi/L_{\perp}$,  $k^{\rm min}_{\parallel} \equiv 2\pi/L_{\parallel}$,
and the volume of the survey is approximately given by $L_\perp^2 L_\parallel$. The lengthscales are defined by:  
\begin{equation}
    L_{\parallel} = \frac{c}{H(z)} \frac{(1+z) B_{\nu, j}}{\nu_{\rm obs, i}}
\end{equation}
where  $B_{\nu,j}$ is the survey bandwidth.  We also have:
\begin{equation}
    L_{\perp}^2 = \chi^2(z) \Omega_{A, j}
\end{equation}
in which $\Omega_{A, j}$ is the solid angle associated with the survey area $S_{A,j}$. This leads to (averaged over the angular variable):
\begin{eqnarray}
   && W_{\rm vol}(k) = 1 - \exp(-k^2/k_{\perp}^2)  \int_0^1 d \mu \exp (k^2 \mu^2 /k_{\perp})^2 \nonumber \\
   &+& \exp(-k^2/k_{\perp}^2) \int_0^1 \exp - \left(k^2 \mu^2 /k_{\parallel}^2 - k^2 \mu^2/k_{\perp}^2\right) d \mu \nonumber \\
    &&  - \int_0^1d \mu  \exp  (- k^2 \mu^2 /k_{\parallel}^2) \nonumber \\
    &\approx & 1 - \frac{\sqrt{ \pi} k_{\parallel}}{2 k}{\rm{Erf}} \left(k / k_{\parallel} \right)
\end{eqnarray}
as found in \citet{hpoiii}.
The full window function associated with the spatial and spectral effects becomes:
\begin{equation}
W_{\rm ij}(k) = W_{\rm beam} (k) W_{\rm vol}(k)
\label{fullwindowfunction}
\end{equation}
Putting it all together, the noise variance of the sub-mm autocorrelation survey is given by:
\begin{equation}
    \text{var}_{\rm ij} = \frac{(P_{\rm i} W_{\rm ij}(k) + P_{\rm N, ij})^2}{{N_{\rm modes, ij}}}
    \label{varianceauto}
\end{equation}
with the number of modes, $N_{\rm modes, ij}$ defined as:
\begin{equation}
    N_{\rm modes, ij} = 2 \pi k^2 \Delta k \frac{V_{\rm surv, ij}}{(2 \pi)^3} 
\end{equation}
in which the $k$ values are equispaced with an interval of $\Delta \log_{10} k = 0.2$,  and the volume of the survey, $V_{\rm surv, ij}$ is given by:
\begin{eqnarray}
V_{\rm surv, ij} &=& 3.7 \times 10^7  {\rm (cMpc}/h)^3 \left (\frac{\lambda_i}{158  \ 
\mu {\rm m}} \right) \left(\frac{1 +z}{8} \right)^{1/2}  \nonumber \\
&& \left(\frac{S_{\rm{A, j}}}{16 {\rm deg}^2} \right) \left(\frac{B_{\nu, j}}{20 
{\rm GHz}} \right)
\label{vsurveyauto}
\end{eqnarray}
The autopower signal and expected noise [with $\sigma_{\rm CII} = (\rm var_{\rm CII})^{1/2}$] in the improved FYST-like  surveys probing [CII]  for $z \sim 5, 6, 7$ are shown in Fig. \ref{fig:autocorrcii}.  For $z \sim 5$ and $z \sim 6$,  the [CII] signal is plotted for both the fits in Table \ref{table:constraints},  denoted as ALPINE-1 and ALPINE-2 in the cases of serendipitous and targeted detections respectively\footnote{The autopower spectra of [OIII] are identical to those in \citet{hpoiii}.}  (which together may be taken to bracket the range allowed by the current observational constraints). For each redshift, the noise measurements are indicated for both the Stage II and Stage III/IV surveys described in Table \ref{table:improved}. For deriving the noise variance from \eq{varianceauto}, we use the ALPINE-1 fit for $P_{\rm CII}$ at $z \sim 5, 6$ and that of the REBELS at $z \sim 7$.

\begin{table*}
\centering
\begin{tabular}{cccccccccc}
\hline
Configuration & $d_{\rm max}$ & $N_{\rm a}$ & $n_{\rm pol}$ & $T_{\rm inst}$[K]  & $A_{\rm eff}$ (m$^2$) & $t_{\rm obs}$[h] &     $S_{\rm A}$[deg.sq] \\ \hline
MWA & 1000 m & 256 &  2 & 28 &  14.5 & 2000 & 1000  \\
SKA-LOW &  40000 m &  512  &  2 & 28  & 964 & 2000  & 1000 \\
\hline
\end{tabular}
\caption{Experimental parameters for the 21 cm surveys. $N_{\rm a}$ here generically denotes the number of independent elements,  i.e.  antennas for the MWA and  SKA-LOW \citep[e.g.,][]{furlanetto2007}.}
\label{table:hi}
\end{table*}

\subsection{21 cm measurements}
The submillimetre configurations considered above are expected to complement  21 cm surveys probing the HI power spectra over the same redshift range. We consider a survey conducted with the Murchison Widefield Array (MWA)\footnote{https://www.mwatelescope.org/telescope/}, which is sensitive to the 21 cm power spectrum over $z \sim  3.7 - 20$, and its successor, the SKA-LOW. The thermal noise in a  interferometric survey is given by \citep[e.g.,][]{bull2014, obuljen2018}:
\begin{equation}
P^{\rm{HI}}_{\rm{N}}(z) = \frac{T^2_\mathrm{sys}(z)\chi^2(z)r_\nu(z) \lambda^4(z)}{A^2_{\rm{eff}} t_{\rm obs} n_{\rm{pol}}n(\textbf{u},z) \nu_{\rm 21}},
\label{noisehiauto}
\end{equation}
typically expressed in units of $\rm{mK}^2 \rm{Mpc}^3$.  

In the above, $A_{\rm eff}$ is the effective collecting area of a single (antenna) element, and  $T_{\rm sys}= T_{\rm sky}+T_{\rm inst}$ is the system temperature, with $T_{\rm sky} = 60{\rm K} \big(300 {\rm MHz}/\nu\big)^{2.25}$ being the sky contribution, and $\nu = 1420 {\rm MHz}/(1+z) \equiv \nu_{21}/(1+z)$ being the observed frequency. The observed wavelength is $\lambda (z) \equiv 21 \ {\rm cm}  (1+z)$, and the parameters $\chi(z)$ and $r_{\nu} (z)$ denote the comoving distance to redshift $z$ and the conversion factor from bandwidth to survey depth\footnote{Note that \eq{noiseautohi} is equivalent to the alternative form employed in the literature \citep[e.g.,][]{mcquinn2006}: 
\begin{equation}
P^{\rm HI}_{\rm{N}}(z) = \frac{T^2_\mathrm{sys}(z)\chi(z)^2(z)\Delta D \lambda^4(z)}{A^2_{\rm{eff}}B t_{\rm obs} n_{\rm{pol}}n(\textbf{u},z)}
\label{noiseautohi}
\end{equation}
which uses the bandwidth, $B_{\nu}$ and the survey depth, $\Delta D \equiv r_{\nu}(z) B/\nu_{21}$ \citep[e.g.,][]{lidz2008, liu2020} assuming that the entire field of view is observed in time $t_{\rm obs}$.} , respectively: 
\begin{equation}
r_{\nu}(z) = \frac{c (1 + z)^2}{H(z)} 
\end{equation}
The baseline density in visibility space is denoted by $n({\textbf{u}}, z)$, which is normalized to the number of independent elements ($N_{\rm a}$). It has the approximate form:
\begin{equation}
n(u, z) = \frac{N_{\rm a}^2}{2 \pi u_{\rm max}^2}
\end{equation}
assuming a constant  density in the $u-v$ plane.  The term $u_{\rm max}$ is, in turn, related to the maximum baseline, $d_{\rm max}$, by
\begin{equation}
u_{\rm max} = \lambda (z)/d_{\rm max}
\end{equation}
The number of independent polarizations is $n_{\rm pol}$, set to 2 for both the surveys under consideration.

For ease of comparison (and cross-correlation) with the sub-millimetre regime, the noise power in \eq{noiseautohi} may be converted to units of (Jy/sr)$^2$  by converting the $T^2_\mathrm{sys}$ term into intensity:
$I_{\rm sys} = {2 k_B T_{\rm sys}}/{\lambda(z)^2}$.

The noise variance of the power spectrum is given by an expression  analogous to that for the sub-millimetre case [e.g., \citet{chen2021}]:
\begin{equation}
{\rm var}_{\rm HI}  = (P_{\rm HI} + P_{\rm N}^{\rm HI})^2/N_{\rm modes}
\end{equation}
in which 
\begin{equation}
N_{\rm modes} = 2 \pi k^2 \Delta k V_{\rm surv}/(2 \pi)^3
\end{equation}
with $V_{\rm surv}$ given by \eq{vsurveyauto}.

We assume a redshift width of $\Delta z \sim 0.3$ at all the redshifts under consideration \citep[e.g.,][]{lidz2008}, which leads to a bandwidth of 
\begin{equation}
B_{\nu } \equiv \Delta z \frac{\nu_{21}}{(1 + z)^2} 
\end{equation}
having values 11.8, 8.7, and 6.7 MHz at $z \sim 5, 6, 7$ respectively.  We also consider the $k$ values to be equispaced in logarithmic space with an interval of $\Delta \log_{10} k = 0.2$.

The parameters for the MWA and SKA-LOW surveys under consideration are provided in Table \ref{table:hi}. 
For the MWA, we adopt the Phase II configuration, with 256 planned antennas of size 4 m $\times$ 4 m and an effective area of $A_{\rm eff} =  14.5$ m$^2$ each,  assumed to be distributed over a super-core inner structure with a maximum baseline $d_{\rm max} = 1$ km \citep{mwa2018, mwa2019}.  For the SKA-LOW, the number of antennas is increased to 512 with effective area $962$ m$^2$ each, distributed\footnote{The parameters are based on  the SKA I Level 0 specifications in https://www.skao.int/sites/default/files/documents/d4-SKA-TEL-SKO-0000007\_SKA1\_Level\_0\_Science\_RequirementsRev02-part-1-signed\_0.pdf} over a maximum baseline of 40 km.  Both surveys cover a field of view of $\sim 1000$ deg$^2$ on the sky.  In both cases, we consider a 2000 h survey,  roughly corresponding to two years of observational time \citep{lidz2008, furlanetto2007, dumitru2019}. 

The auto-correlation signal and expected noise [with $\sigma_{\rm HI} = (\rm var_{\rm HI})^{1/2}$]  power for MWA and SKA-LOW for each of the three redshifts under consideration are plotted in Fig. \ref{fig:autocorrhi}. { We follow previous work \citep{moriwaki2019, dumitru2018} in assuming that all modes above $k \sim 0.1 h$ Mpc$^{-1}$  -- which represents the limiting scale above which foregrounds could be expected to influence the signal \citep[e.g.][]{lidz2011} -- are available from the 21 cm power spectrum.}

\begin{figure}
    \centering
    \includegraphics[width = 0.9\columnwidth]{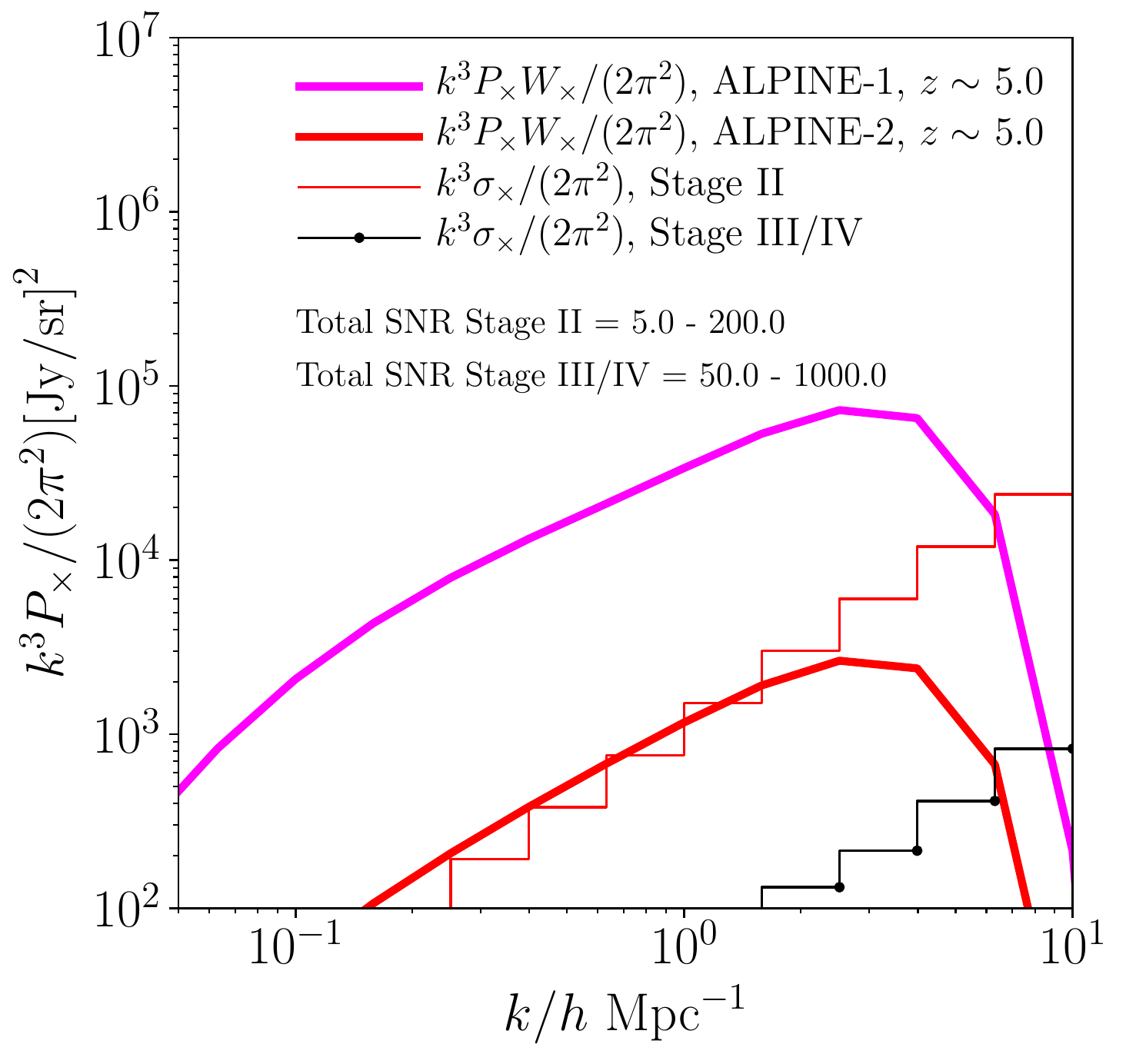} \includegraphics[width = 0.9\columnwidth]{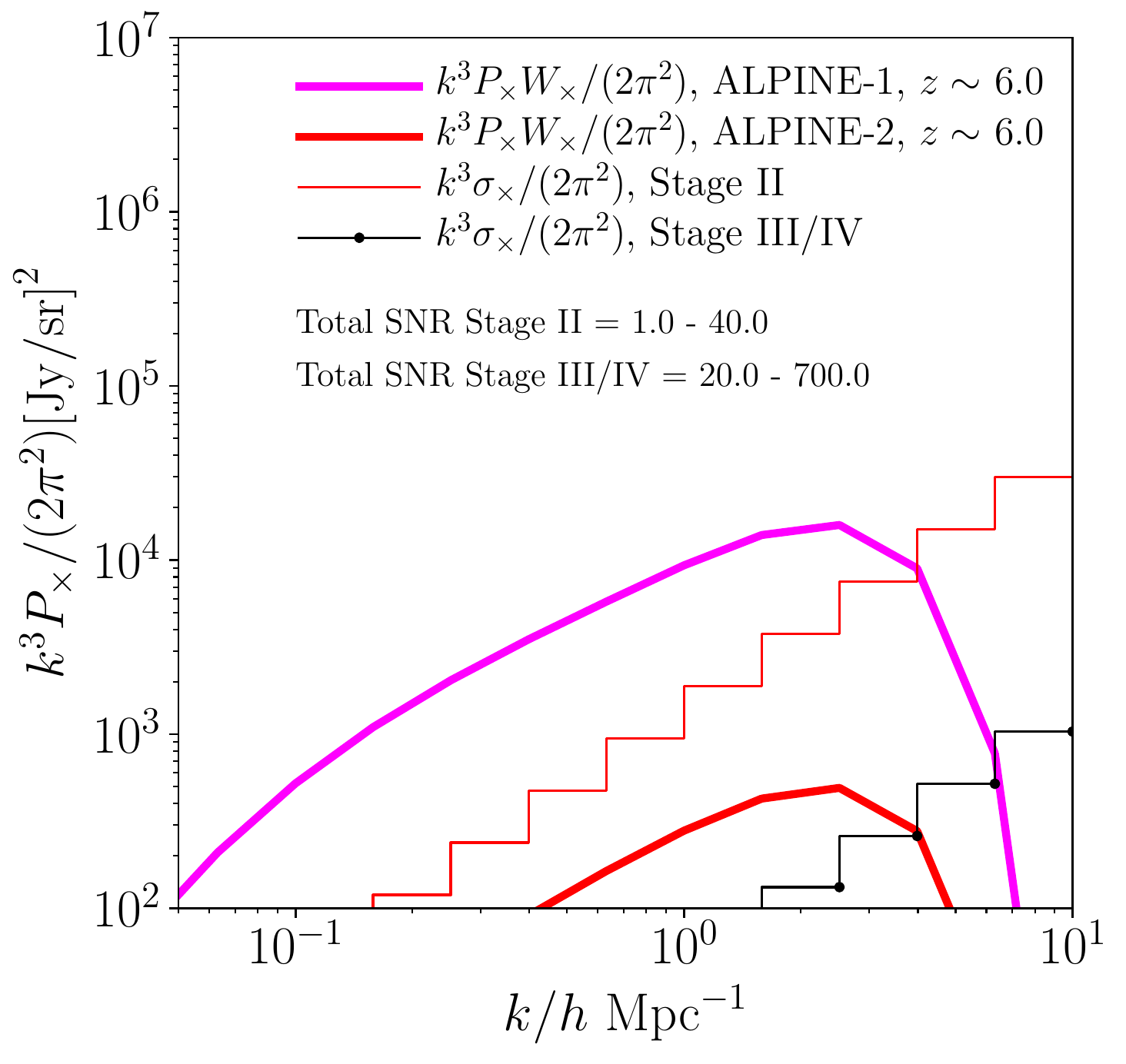}\\
    \includegraphics[width = 0.9\columnwidth]{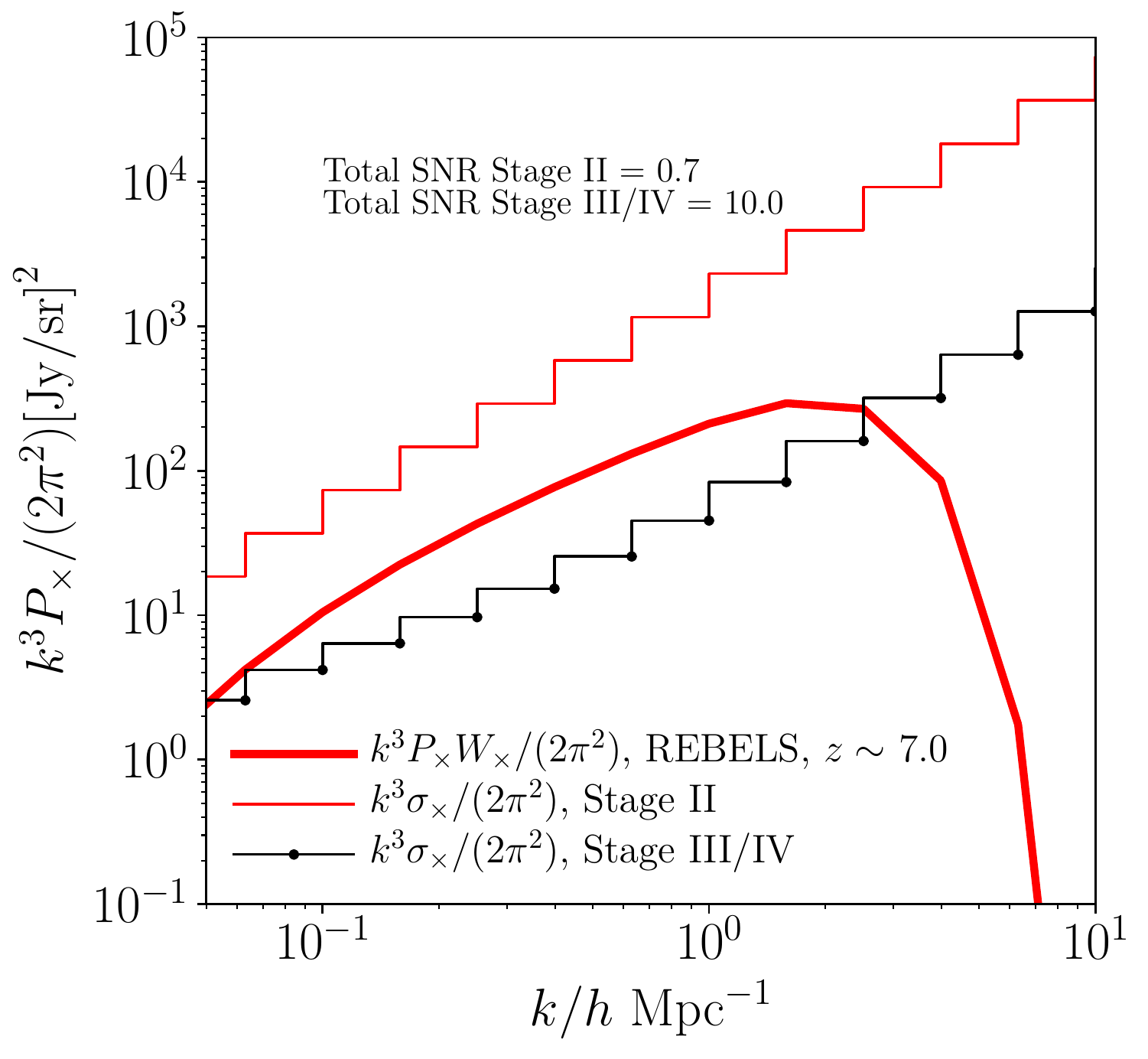}
    \caption{Cross-correlation signal (thick line) and noise (thin steps) power  for the MWA and FYST-like experiment combination { with parameters as in Table \ref{table:improved}}, probing the 21 cm - [CII] power at $z \sim 5, 6$ and 7.  As in Fig. \ref{fig:autocorrcii}, the [CII] signal power is calculated for both the ALPINE-1 and ALPINE-2 based fitting forms at $z \sim 5,6$, and for that based on the REBELS sample at $z \sim 7$. The total SNR [obtained by using \eq{snrcross}] is marked on each of the figures.}
    \label{fig:crosspowermwafyst}
\end{figure}

\begin{figure}
    \centering
    \includegraphics[width = 0.9\columnwidth]{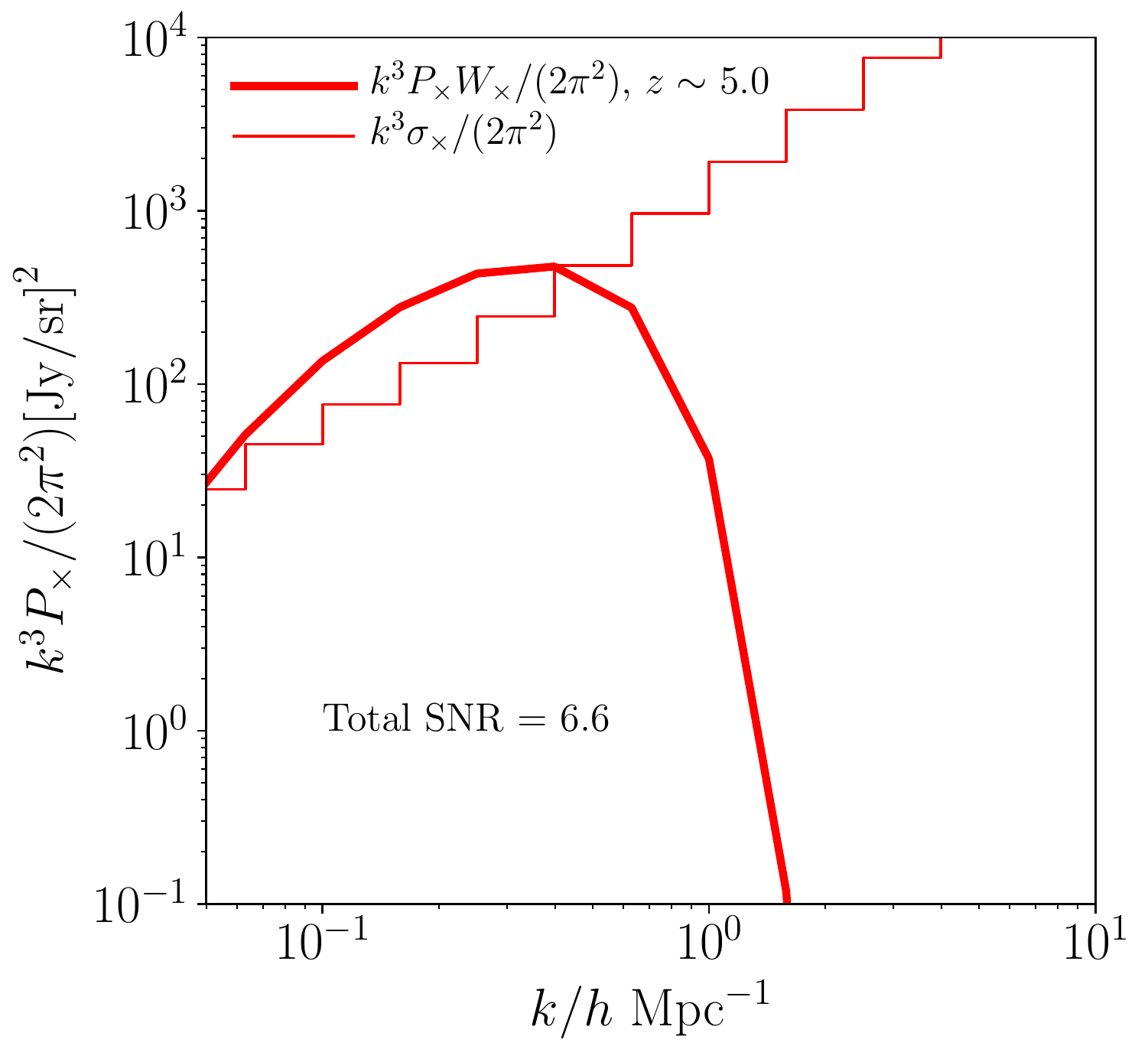} \includegraphics[width = 0.9\columnwidth]{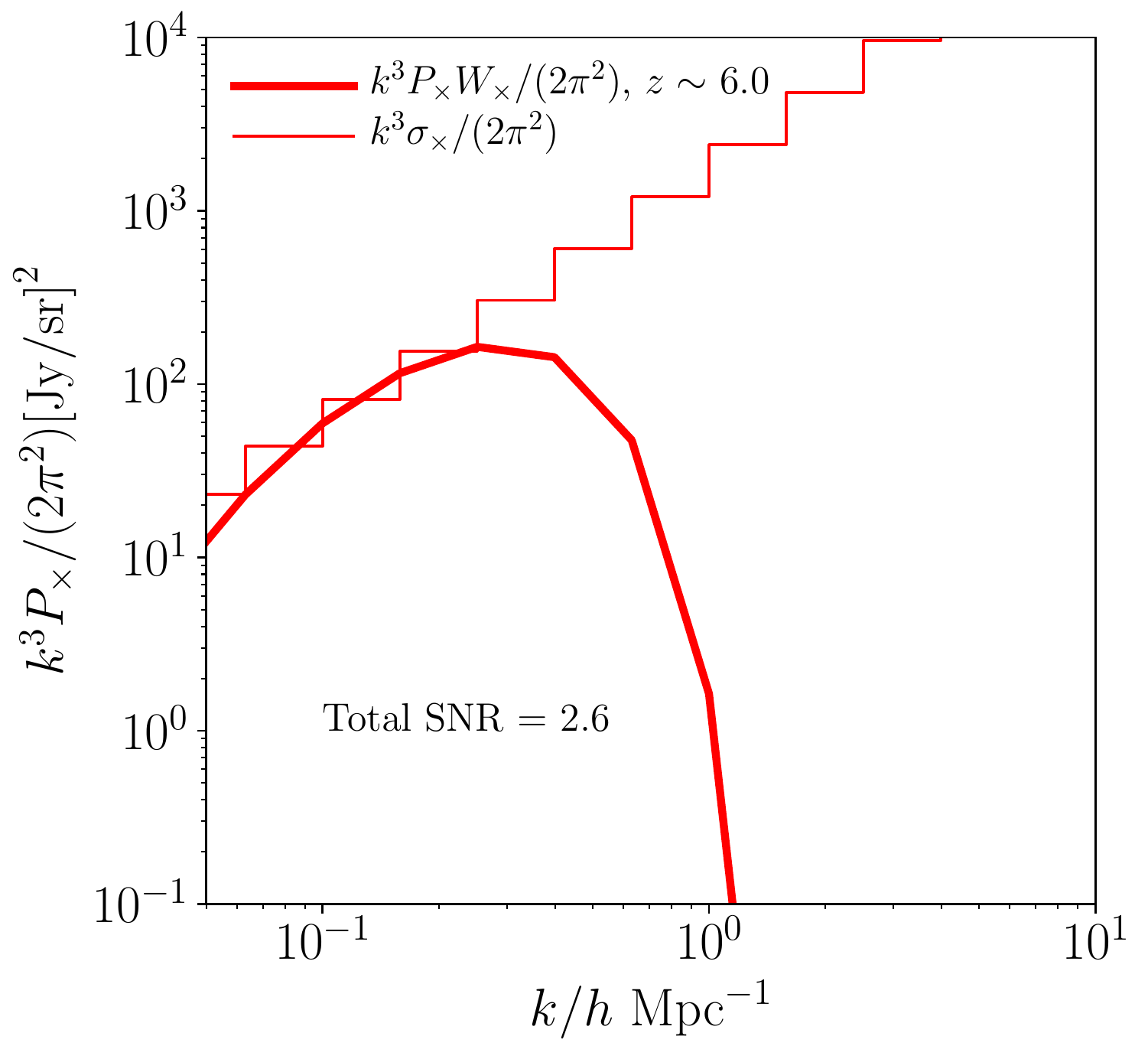} \\
    \includegraphics[width = 0.9\columnwidth]{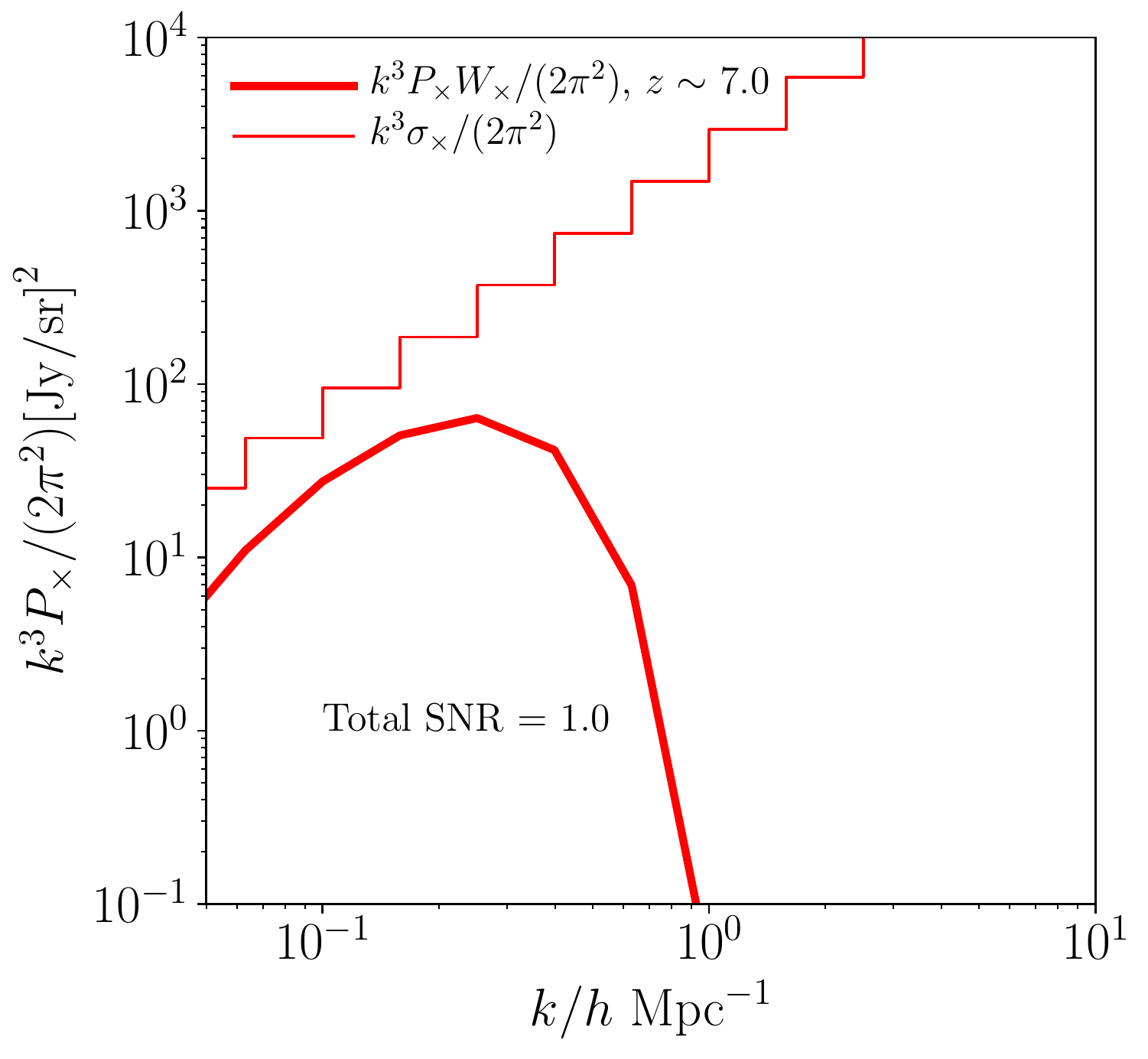}
    \caption{Cross-correlation signal (thick line) and noise (thin steps) power for the MWA and EXCLAIM-like experiment combination { with parameters as in Table \ref{table:improved}}, probing the 21 cm - [OIII] power at $z \sim 5, 6 , 7$. The total SNR (obtained by using \eq{snrcross}) is marked on each of the figures.}
    \label{fig:crosspowermwaexclaim}
\end{figure}

\section{Cross-correlating an [HI] and a sub-mm survey}
\label{sec:crosscorr}
We now forecast the cross-correlation power and associated noise measured by a 21 cm - submillimetre survey combination over the redshift range $z \sim 5-7$.
Given the individual power spectra in \eq{powerspechi} and \eq{powerspeccii}, the cross-correlation signal power at a given redshift can be approximated  as:
\begin{equation}
    P_{\times} = (P_{\rm CII/OIII} P_{\rm HI})^{1/2}
    \label{crosscorrpow}
\end{equation}
The above expression is equivalent, modulo a shot-noise like contribution, to that obtained from averaging the HI masses/luminosities in Eqs. (\ref{twohalo}) and (\ref{CIIspint}) respectively \citep[e.g.,][]{liu2021, beane2018, hpoiii, oxholm2021}. This term is, however, expected to be subdominant on clustering scales \citep[e.g.,][]{liu2021} from where the bulk of the cross-correlation signal originates. We follow  \citet{breysse2022,hpoiii} in assuming that both intensity maps follow Gaussian statistics at the scales of interest; this is also expected to hold in the large-scale regime which is not suppressed by the beam. 

We use the specifications of the  surveys discussed in the previous section $z \sim 5, 6, 7$ and listed in Tables \ref{table:improved}  and \ref{table:hi}, assuming fiducial observing times of 4000 h, 4000 h and 2000 h respectively for the improved FYST-like, EXCLAIM-like and MWA surveys, and complete area overlap between them.

The noise in the cross-correlation of [CII] and [HI] surveys is calculated by defining the total number of modes:
\begin{equation}
N_{\rm modes, \times} = 2 \pi k^2 \Delta k \frac{V_{\rm surv, \times}}{(2 \pi)^3}
\label{nmodes}
\end{equation}
where the $k$ values are equispaced with an interval of $\Delta \log_{10} k = 0.2$. The volume relevant to the cross-correlation  is:
 \begin{eqnarray}
V_{\rm surv, \times} &=& 3.7 \times 10^7  {\rm (cMpc}/h)^3 \left (\frac{\lambda}{158  \ 
\mu {\rm m}} \right) \left(\frac{1 +z}{8} \right)^{1/2}  \nonumber \\
&& \left(\frac{S_{\rm{A, \times}}}{16 {\rm deg}^2} \right) \left(\frac{B_{\nu}}{20 
{\rm GHz}} \right) \, ,
\label{vsurveycross}
\end{eqnarray}
 calculated by using the area of the smaller volume survey (which corresponds to the sub-millimetre survey in all cases considered here) as $S_{A, \times}$.

\begin{figure}
    \centering
    \includegraphics[width = 0.9\columnwidth]{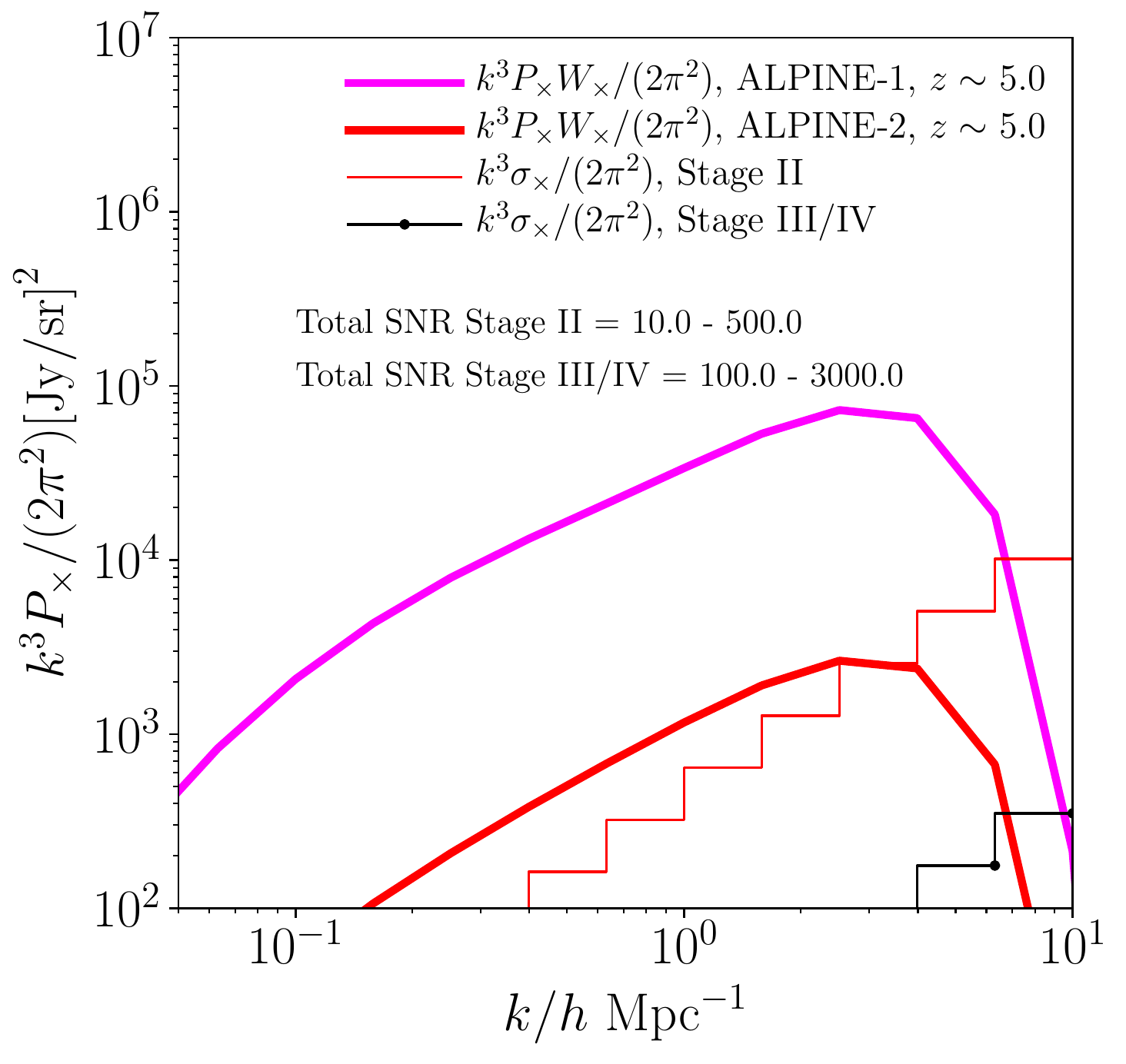} \includegraphics[width = 0.9\columnwidth]{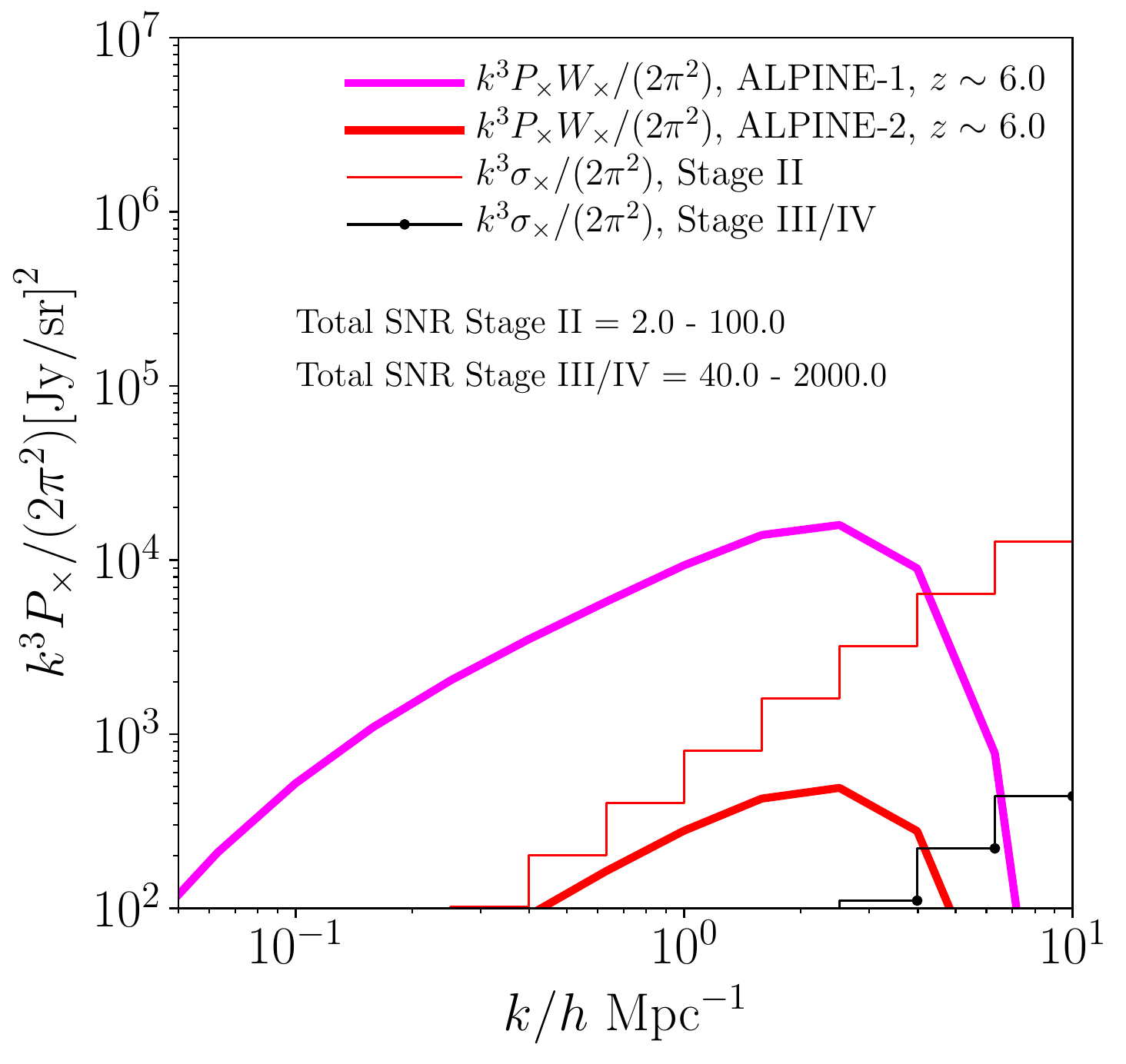}\\
    \includegraphics[width = 0.9\columnwidth]{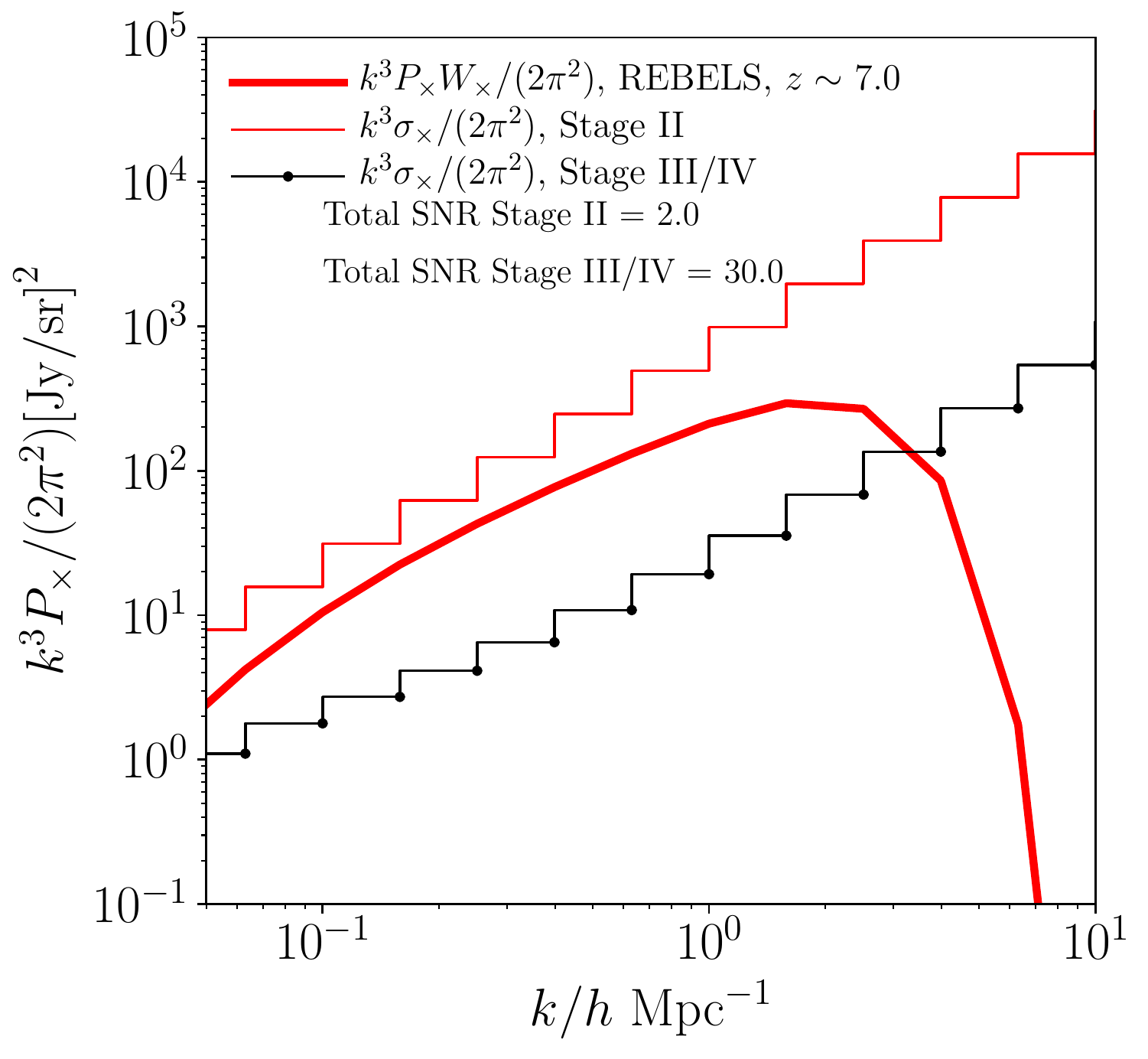}
    \caption{Same as Fig. \ref{fig:crosspowermwafyst}, but for the SKA-LOW and FYST-like experiment combination.}
    \label{fig:crosspowerskafyst}
\end{figure}

\begin{figure}
    \centering
    \includegraphics[width = 0.9\columnwidth]{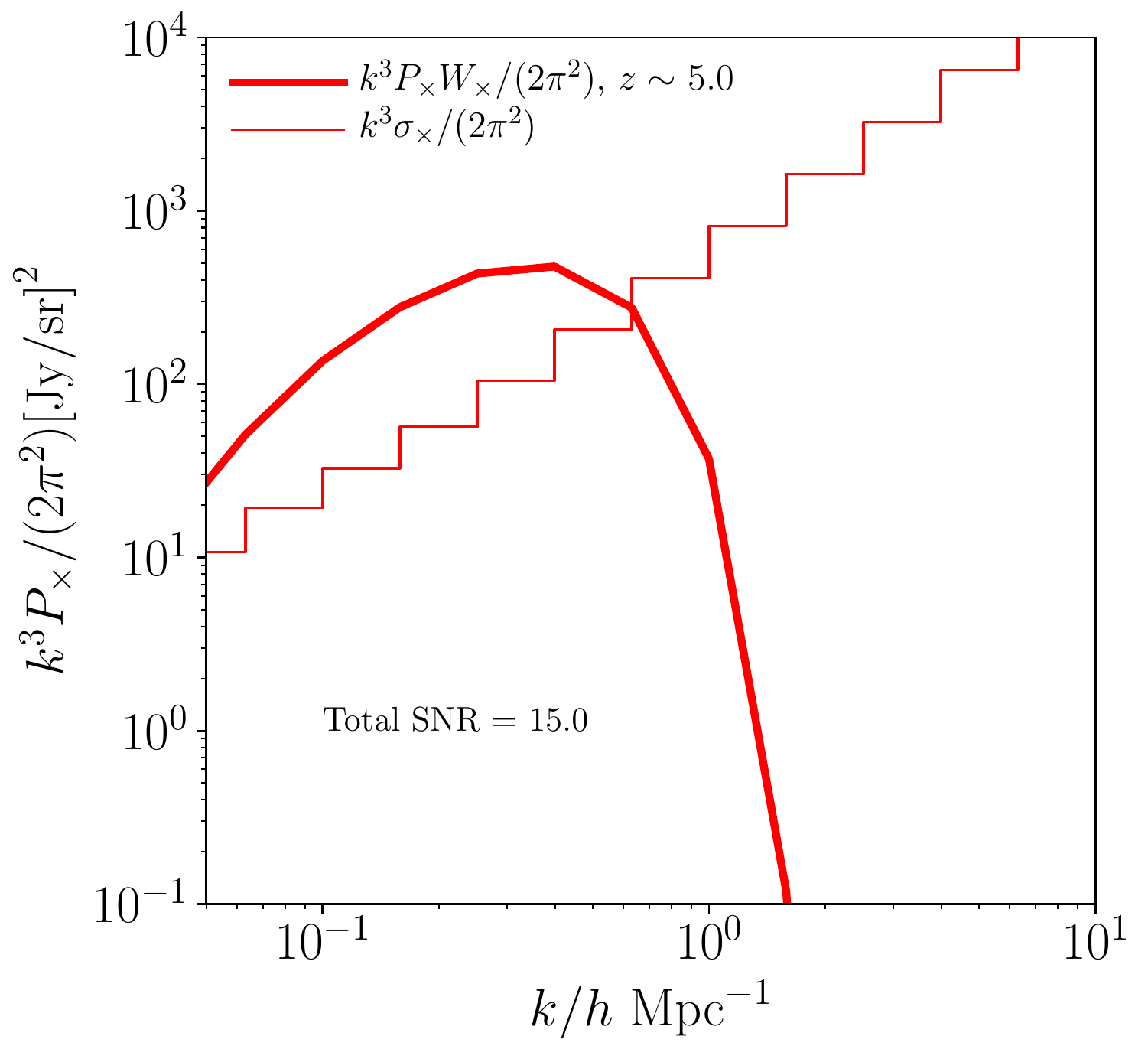} \includegraphics[width = 0.9\columnwidth]{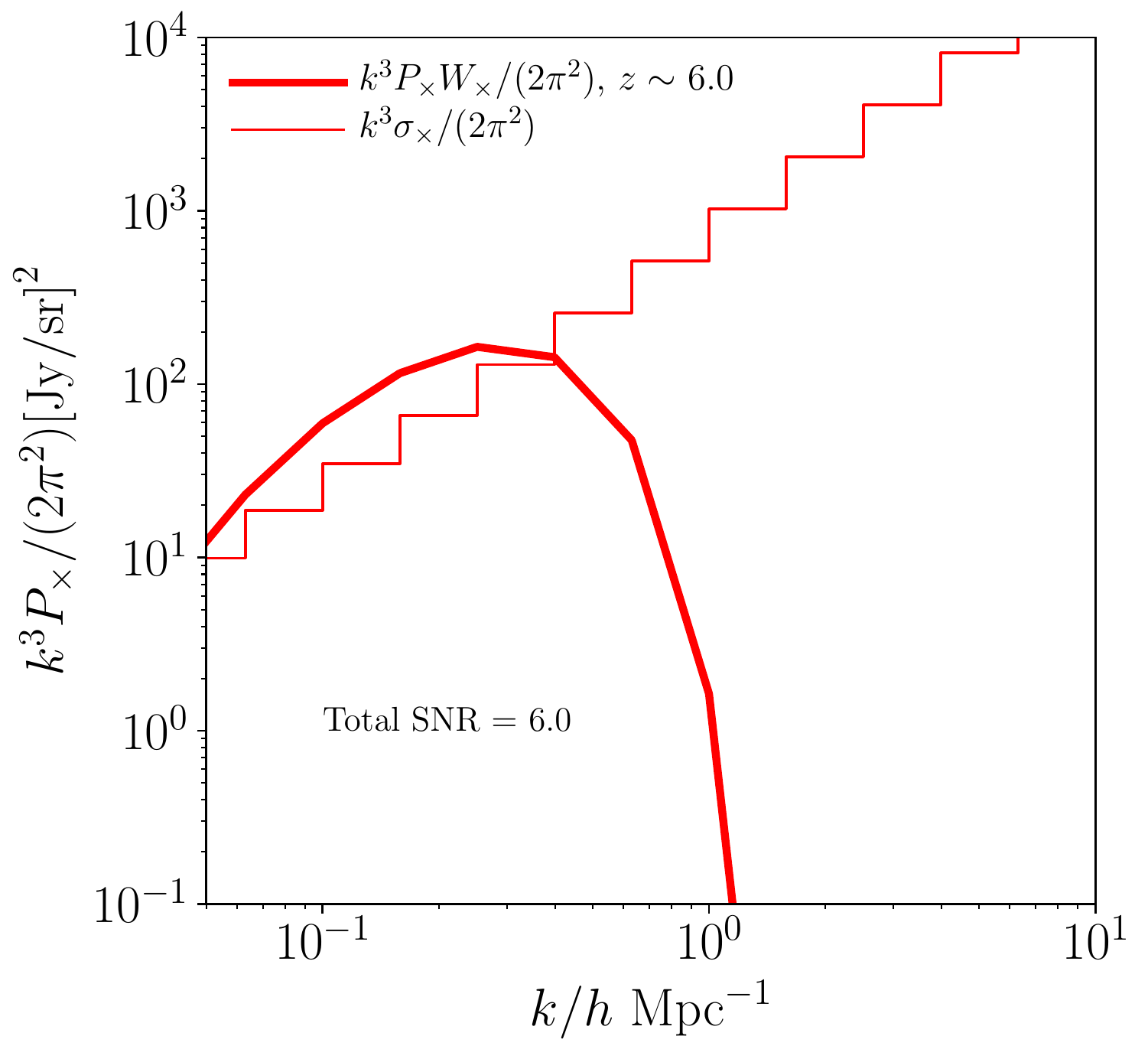}\\
    \includegraphics[width = 0.9\columnwidth]{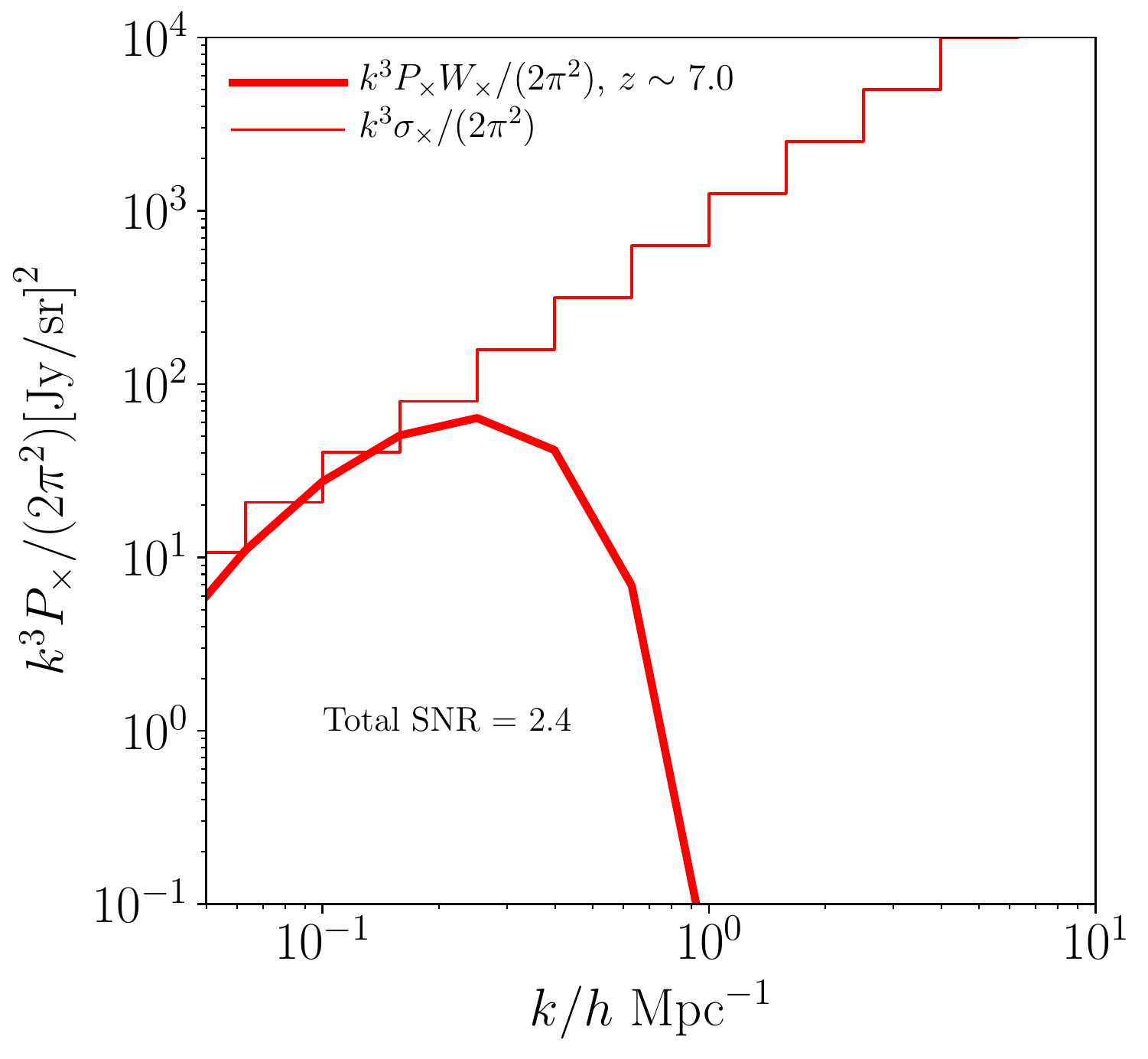}
    \caption{Same as Fig. \ref{fig:crosspowermwaexclaim}, but for the SKA-LOW and EXCLAIM-like experiment combination.}
    \label{fig:crosspowerskaexclaim}
\end{figure}

Accounting  for the finite spatial and spectral resolution of the submillimetre configurations, { as well as the volume effects} described in Sec. \ref{sec:volres}, the variance of the cross-correlation survey is given by \citep[e.g.,][]{hpoiii}:
\begin{eqnarray}
    {\rm var}_{\times}(k) &=& \left((P_{\rm CII/OIII}W_{\rm CII/OIII}(k) + P_{\rm N(CII/OIII)}) \right. \nonumber\\
   & & \left. (P_{\rm HI} + P^{\rm N}_{\rm HI}) \right. \nonumber\\
    &+& \left. P_{\times}^2 W_{\rm CII/OIII}(k) \right)/2 N_{\rm modes, \times}
\label{varciicross}
\end{eqnarray}
in which $W_{\rm CII/OIII}, P^{\rm N}_{\rm HI}$  and $P_{\rm N(CII/OIII)}$ follow \eq{fullwindowfunction}, \eq{noisehiauto} and \eq{noiseciiauto} respectively. Given the signal and noise variance, the signal-to-noise ratio (SNR) of the measurement is calculated as:
\begin{equation}
{\rm{SNR}} = \left(\sum_k \frac{P_{\times}^2(k) W_{\rm CII/OIII}(k)}{{\rm{var}}_{\times}(k)}\right)^{1/2} \, ,
    \label{snrcross}
\end{equation}
The cross-correlation signal (modulated by $W_{\times} \equiv W_{\rm CII/OIII}$) and noise [$\sigma_{\times} = (\rm var_{\times})^{1/2}$] power for the improved FYST-like (both the Stage II Stage III/IV configurations in Table \ref{table:improved}) and  MWA experiment combination is plotted in Fig. \ref{fig:crosspowermwafyst}.  We use \eq{crosscorrpow} for the signal,\footnote{We are chiefly interested in the magnitude of the signal and do not model the scale at which the cross-power is expected to transition from positive to negative values, which in turn depends on the evolution of the bubble size \citep[e.g.,][]{lidz2011,dumitru2019} and is presently unconstrained by the data. In some reionization scenarios, this scale may be close to where the beam effect suppresses the signal power in Figs. \ref{fig:crosspowermwafyst} - \ref{fig:crosspowerskaexclaim}.}  with $P_{\rm CII}$ following both the fits in Table \ref{table:constraints} [denoted by ALPINE-1 and ALPINE-2 for the serendipitous and targeted detections respectively]. The survey volume in \eq{vsurveycross} is that of the [CII] experiment, which is the smaller survey. For calculating the noise variance, we use
 \eq{varciicross}, in which $P_{\rm CII}$ corresponds to the ALPINE-1  fit for both $z \sim 5, 6$ and to the REBELS fit for $z \sim 7$. The $P_{\rm N(CII/OIII)}$ term in \eq{varciicross} is calculated for both the Stage II and the Stage III/IV configurations considered in Sec. \ref{sec:formalism}. The total signal-to-noise ratio [obtained by using \eq{snrcross}] is indicated on each figure. For $z \sim 5$ and $z \sim 6$, we show the range in the SNR bracketed by the ALPINE-1 and ALPINE-2 results for each of the two experimental configurations. It can be seen that 
a detection of the signal is possible in all cases with this configuration, with the maximum SNR reaching values of a few ten to a few hundred depending on the redshift. 

Fig. \ref{fig:crosspowermwaexclaim} shows the cross-correlation signal and noise power at $z \sim 5,6$ and 7 for the MWA and EXCLAIM-like experiment probing [OIII], with the [OIII] signal power calculated using the fitting function in \eq{oiiilumfunc}. The noise variance is again calculated using \eq{varciicross}, in which  $P_{\rm N(CII/OIII)}$ follows \eq{noiseciiauto} with the parameters of the [OIII] survey described in Table \ref{table:improved}. The number of modes is calculated using the parameters of the EXCLAIM-like experiment, which is the smaller survey. Again, the signal is detectable in all cases with this experiment combination, with the SNR reaching values of a few.

Finally, Figs. \ref{fig:crosspowerskafyst} and \ref{fig:crosspowerskaexclaim} show the signal and noise forecasts for a SKA-LOW survey cross-correlated with the improved FYST-like and the EXCLAIM-like configurations over the same redshift range. Compared to the previous results, it can be seen that the forecasts improve by a factor of a few to an order of magnitude, with secure detections expected at all redshifts in both the [CII]-HI and [OIII]-HI cases.

\section{Discussion}
\label{sec:conclusions}
We have explored the synergies between 21 cm surveys  and sub-millimetre intensity mapping experiments probing the mid-to end stages of reionization ($z \sim 5-7$). Using the latest findings of the ALMA (ALPINE and REBELS) surveys at $z \sim 4.5 - 7$, we developed functional forms for the evolution of the [CII] luminosity to host halo mass relation at these epochs, finding the best-fitting parameters to be in good agreement with the results of previous observations and simulations (with the ALPINE targeted and serendipitous detections bracketing the allowed range in [CII] luminosities over $z \sim 4.5 - 6$). { This follows the empirical approach to modelling the tracer - halo mass relation from the data alone \citep[e.g.,][]{hpcii2019, hpoiii}, which is complementary to forward modelling approaches that start from the physics of [CII] to arrive at the expected intensity of emission \citep[e.g.,][]{gong2012}.} We have considered four different experimental configurations measuring the cross-correlation signal: between improved versions of the FYST-like and EXCLAIM-like experiments introduced in \citet{hpoiii}, and an MWA and SKA-LOW survey probing the 21 cm power spectrum at the same redshifts.
The cross-correlation can improve the detection significance by factors of a few to a few ten beyond that from the individual sub-millimetre and 21 cm forecasts. The Stage II improvements in the FYST-like configuration should already be able to detect the cross-correlation power with a good significance at $z \sim 5 -7$, with the signal-to-noise going up to a few hundreds  for the optimistic end of the [CII] luminosity function. With SKA-LOW, a secure detection is expected in all cases out to $z \sim 7$.  The effects of foreground contamination and instrumental response on the 21 cm signal are expected to be sub-dominant in the cross-correlation measurements, especially for the supercore configurations of  the MWA and SKA-LOW.

{ The effects of possible interloper contamination on the sub-millimetre signals (such as from high-J CO transitions)  are expected to have a negligible effect on the cross-power measurement.  While  avoiding the low $k_\parallel$ modes in the data may be needed for mitigating the spectrally smooth continuum emissions from Galactic dust and the cosmic infrared background, this is not expected to degrade the SNR significantly \citep{dizgah2018}.  The  approach of targeted voxel masking to reduce CO contamination around infrared-luminous galaxies  has already been found to reduce the CO fluctuation level to less than 10\% of
the CII power (at the representative wavenumber of $k \sim 0.1 h \ {\rm Mpc}^{-1}$),
while only removing 8\% of the survey volume \citep{sun2018}. Applying such a targeted masking across the large fields considered here with a similar level of success would lead to little impact
on our SNR forecasts. Covariance-based filtering \citep{chung2023}, overlap with known tracers like Ly-$\alpha$ in the same field and using the anisotropic power spectrum \citep{lidz2016}, would,  in addition to targeted masking, lead to a fairly robust mitigation of the interloper foregrounds. A possible bias due to interlopers  is also not expected to influence the cross-correlation with 21 cm, since any interloping lines for [CII]/[OIII] and 21 cm are widely enough separated in redshift so as to be statistically independent and uncorrelated.}

Given the current uncertainties in the high redshift observations (both in HI and the sub-mm lines), we do not attempt a full parameter estimation process at this stage, except to note that cross-correlating intensity maps of 21 cm and sub-mm surveys promises stringent constraints on the evolution of the baryon cycle during reionization.  The [CII] observations at high and low redshifts  can be combined to study the evolution of the [CII] luminosity - halo mass relation over cosmic time and its consequences for  the global star-formation history, which we leave to future work.

We have focused chiefly on the magnitude of the 21 cm auto- and cross-correlation power -- rather than the details of its scale dependence, which are presently  poorly constrained. To this end, we have extended the model of HI occupation in dark matter haloes out to $z \sim 7$, in line with the mild evolution of total HI density  expected across these epochs \citep{peroux2020, heintz2022}. While the results are expected to be robust around $z \sim 5$, this is less clear at $z \sim 6-7$, where the halo model is likely conservative since the diffuse HI in the high-redshift IGM is expected to make a larger contribution to the cross-correlation signal than we assume here.  To make a rough estimate of this effect, we consider the expected increase in the overall signal-to-noise ratios for an assumed neutral fraction of $x_{\rm HI} \sim 0.5 - 0.8$ at $z \sim 6 - 7$ respectively, as suggested by recent analyses \citep{mason2018, hoag2019} of Lyman Break Galaxies at these epochs. The HI power is given by $\Delta_{21}^2 (k) \approx x_{\rm HI}^2 T_0^2 \Delta^2_{\rm dm} (k)$, where $T_0^2 \sim (28 \ {\rm mK})^2 (1 +z)/10$ and $\Delta^2_{\rm dm} (k)$ is the dimensionless dark matter power spectrum. In models such as \citet{lidz2008}, the above range of neutral hydrogen fractions leads to $\Delta_{21}^2 \approx 0.5 - 8$ Jy/sr on scales $k \sim 0.1 h$ Mpc$^{-1}$ (which is assumed to be much larger than the bubble size). Thus, such a scenario may boost the the auto-power in Fig. \ref{fig:autocorrhi} by  a factor of about 5 - 80 on these scales, and the cross-correlation forecasts by up to an order of magnitude.

Future observations (also of the global 21 cm signal) will allow for the possibility of modifying the HI-halo mass relation to take into account the detailed physics associated with reionization and bubble overlap, while bringing in new parameters into the problem (such as the IGM clumping factor, the escape fraction of ionizing photons and feedback from winds and supernovae). Recent simulation-based approaches have indicated good prospects for constraining these parameters in fiducial scenarios of the reionization process \citep{dumitru2019, lidz2008}. Spatial effects that influence the cross-correlation signal will also become relevant at this stage, such as the anticorrelation expected between HI and [CII]/[OIII] above the characteristic scale of ionized regions \citep{lidz2011} -- which, if measurable, is also  an important diagnostic of the contribution of low- and high-mass sources to reionization \citep{dumitru2019}.

The high signal-to-noise ratios promise an optimistic outlook for probing the detailed physics at different stages of reionization using future synergies between 21 cm and sub-mm surveys. These would be a useful complement to deep, targeted surveys of galaxies in this era, conducted e.g., with the \textit{James Webb Space Telescope} (JWST) out to $z \sim 9$ and beyond. Adding information from complementary line tracers, such as the carbon monoxide (CO),  Ly$\alpha$, H$\alpha$, [OI] and [NII] lines observed with the CO Mapping Array Project - Epoch of Reionization Array \citep[COMAP-ERA;][]{breysse2022}, the Cosmic Dawn Intensity Mapper \citep[CDIM;][]{cooray2016, cooray2019}, and the  Spectro-Photometer for the History of the Universe, EoR, and Ices Explorer \citep[SPHEREx;][]{dore2014}, would open up further  opportunities to precisely trace the evolution of this exciting period.

\section*{Acknowledgements}
I thank Miroslava Dessauges,  Pascal Oesch and Omkar Bait for useful discussions, Adam Lidz and Cathryn Trott for detailed and helpful comments on the manuscript,  Pascal Oesch for sharing the data from the targeted ALPINE and REBELS surveys {and the referee for a helpful report}.  My research is supported by the Swiss National Science Foundation via Ambizione Grant PZ00P2\_179934.

\section*{Data availability}
No new data were generated in support of this research. The software underlying this article will be shared on reasonable request to the author.

\bibliographystyle{mnras}
\bibliography{mybib}

\bsp
\label{lastpage}
\end{document}